\documentclass[lettersize,journal]{IEEEtran}
\usepackage{amsmath,amssymb,amsfonts}
\newtheorem{claim}{Claim}
\newtheorem{theorem}{Theorem}
\newtheorem{corollary}{Corollary}

\newtheorem{proposition}{Proposition}
\newtheorem{lemma}{Lemma}
\newtheorem{assu}{Assumption}
\newtheorem{pb}{Problem}
\newtheorem{remark}{Remark}
\usepackage{tabularx}
\usepackage{todonotes}
\usepackage{cite}
\usepackage{algorithmic}
\usepackage{algorithm}
\def\BibTeX{{\rm B\kern-.05em{\sc i\kern-.025em b}\kern-.08emT\kern-.1667em\lower.7ex\hbox{E}\kern-.125emX}}
\usepackage{url}
\allowdisplaybreaks[4]
\usepackage{balance}
\usepackage[caption=false,font=footnotesize,labelfont=rm,textfont=rm]{subfig} 

\ifCLASSINFOpdf
\usepackage{graphicx}

\else

\fi

\begin{document}
	
	\title{Age Optimal Sampling for Unreliable Channels under Unknown Channel Statistics}
\author{Hongyi~He,~\IEEEmembership{Student Member,~IEEE},~Haoyue~Tang,~\IEEEmembership{Member,~IEEE},~Jiayu~Pan,~Jintao Wang,~\IEEEmembership{Senior Member,~IEEE}, \\Jian Song,~\IEEEmembership{Fellow,~IEEE}, and Leandros~Tassiulas,~\IEEEmembership{Fellow,~IEEE}
	\thanks{H.~He, J.~Wang are with the Department of Electronic Engineering and Beijing National Research Center for Information Science and Technology, Tsinghua University, Beijing 100084, China (e-mail: {hehy22@mails.tsinghua.edu.cn; wangjintao@tsinghua.edu.cn}).}
	\thanks{H.~Tang was with Yale University. She is now with Meta Platforms, Inc. (e-mail: tanghaoyue13@tsinghua.org.cn).}
    \thanks{J.~Pan is with the School of Software Technology, Zhejiang University, Zhejiang, China (e-mail: jiayupan26@zju.edu.cn).}
	\thanks{J.~Song is with the Department of Electronic Engineering, Tsinghua University, Beijing 100084, China, and also with Shenzhen International Graduate School, Tsinghua University, Shenzhen 518055, China (e-mail: jsong@tsinghua.edu.cn).}
	\thanks{L.~Tassiulas is with the Department of Electronic Engineering, Yale University, New Haven, USA (e-mail: leandros.tassiulas@yale.edu).}}
	
	\maketitle
	

	\begin{abstract}
    In this paper, we study a system in which a sensor forwards status updates to a receiver through an error-prone channel, while the receiver sends the transmission results back to the sensor via a reliable channel. Both channels are subject to random delays. To evaluate the timeliness of the status information at the receiver, we use the Age of Information (AoI) metric. The objective is to design a sampling policy that minimizes the expected time-average AoI, even when the channel statistics (e.g., delay distributions) are unknown. We first review the threshold structure of the optimal offline policy under known channel statistics and then reformulate the design of the online algorithm as a stochastic approximation problem. We propose a Robbins-Monro algorithm to solve this problem and demonstrate that the optimal threshold can be approximated almost surely. 
    Moreover, we prove that the cumulative AoI regret of the online algorithm increases with rate $\mathcal{O}(\ln K)$, where $K$ is the number of successful transmissions. In addition, our algorithm is shown to be minimax order optimal, in the sense that for any online learning algorithm, the cumulative AoI regret up to the $K$-th successful transmissions grows with the rate at least $\Omega(\ln K)$ in the worst case delay distribution. Finally, we improve the stability of the proposed online learning algorithm through a momentum-based stochastic gradient descent algorithm. Simulation results validate the performance of our proposed algorithm.
	\end{abstract}


	\IEEEpeerreviewmaketitle
	\section{Introduction}

    \IEEEPARstart{T}{he} proliferation of real-time control systems, such as autonomous driving, industrial automation, and health monitoring, has created increasing demands for timely status updates to ensure effective monitoring and control \cite{yates2021age, xie2024scheduling}. To measure the freshness of the status update, a new Quality of Service (QoS) metric, the Age of Information (AoI), has been proposed \cite{kaul2012realtime}. By definition, the Age of Information is the time difference between the current time and the generation time of the freshest status update stored at the receiver. A smaller AoI indicates that the status information at the receiver is more up-to-date, enabling faster and more informed decision-making.

    Designing sampling policies to minimize the AoI performance has received significant attention \cite{wang2019when, tang2023agea, xiao2024adaptive, zhou2019joint, tang2020minimizing, abd-elmagid2020reinforcement, ceran2021reinforcement, pan2023optimal, bedewy2021optimal, yates2015lazy, pan2023ageoptimal, sun2017update}. 
    For point-to-point communication channel with a random delay, it is shown that the naive ``\emph{zero-wait}'' sampling policy, i.e., take a new sample immediately once the previous sample has been received, is not AoI minimum. To minimize the average AoI performance, the sampler should take a new sample if the information stored at the receiver becomes stale, i.e., when the AoI exceeds a certain threshold  \cite{yates2015lazy}. Finding the optimum sampling threshold for channels with reliable transmission and instantaneous feedback has been investigated in \cite{sun2017update}. Moreover, considering that the backward channel is non-ideal in practical communication systems, the authors in \cite{tsai2020ageofinformationa} introduced a two-way delay model and derived the optimal sampling policy. Furthermore, recent work has considered unreliable transmissions with two-way delay and derived age-optimal sampling policies that adapt to such conditions \cite{pan2023ageoptimal}. These optimal policies typically exhibit a threshold-based structure, and the computation of the optimal threshold requires that the closed-form expression of the transmission statistics, such as the delay distribution, are known in advance.

   When the channel statistics are unknown, online learning can provide provable and low computational complexity algorithms that can learn the optimal threshold adaptively \cite{atay2021aging, banerjee2020fundamental, tripathi2021online, tang2023age, tsai2022distributionoblivious, xu2022schedule}. Online policies for reliable channels have been proposed in \cite{banerjee2020fundamental, tripathi2021online, tang2023age, tsai2022distributionoblivious, xu2022schedule}. Tang {\emph{et al}}. used Robbins-Monro algorithm to obtain the age-optimal sampling policy adaptively for a one-way delay model \cite{tang2023age, tang2022sendinga}. Specifically, the almost sure convergence properties of the average AoI performance are verified through the stochastic differential equations (SDE). Furthermore, a similar online algorithm is derived to minimize the MSE when sampling a wiener process in \cite{tang2022samplinga, tang2023sampling}. However, these studies \cite{tang2023age, tang2022sendinga, tang2022samplinga, tang2023sampling} assume reliable transmissions. In \cite{tsai2020ageofinformationa, tsai2022distributionoblivious}, the authors proposed an online algorithm for sampling in a status update system, where both the forward and backward links have non-zero delay. However, theoretical analysis, such as the convergence rate of the optimality gap, i.e., the cumulative AoI difference between the proposed algorithm and the optimal offline policy, and the worst-case lower bound for the optimality gap, is not provided in \cite{tsai2020ageofinformationa, tsai2022distributionoblivious}. In addition, these studies \cite{tsai2020ageofinformationa, tsai2022distributionoblivious} did not take into account unreliable transmissions as well. To the best of our knowledge, provable online learning algorithms for sampling systems with two-way delay and unreliable transmissions are still lacking.

	Moreover, the value of the sampling threshold learned through the vanilla Robbins-Monro algorithm oscillates when the transmission delay distribution has a high variance.
Therefore, modifications to the above algorithms are needed to mitigate the variance brought by delay randomness. Variance reduction techniques, ranging from averaged-gradient to momentum acceleration \cite{johnson2013accelerating, defazio2014saga, nguyen2017sarah, Cutkosky2019momentum, liu2020improved} can accelerate convergence. Among them, momentum-based methods utilize past gradients or sample information to alleviate the randomness of the current sample and accelerate the convergence without a large computational burden \cite{Cutkosky2019momentum, liu2020improved}. The successful applications of the momentum-based method inspire us to apply it to the online learning algorithm.

   Motivated by the previously mentioned challenges and research gaps, in this paper, we aim to minimize AoI with unknown channel statistics under one of the most general channel settings in the literature: unreliable transmissions with random two-way delay. Note that the above channel settings are similar to that of \cite{pan2023optimal}, but without access to channel statistics. The
   theoretical framework in this work is most relevant to \cite{tang2023age} but
   with significant differences. Due to unreliable transmissions, we modify the existing offline optimal policy to make the online algorithm effective. Additionally, we construct different worst-case distributions to prove the minimax error bound.
    The main contributions of this work are as follows:
     \begin{itemize}
         \item We reformulate the age-optimal sampling problem under unreliable transmissions into a stochastic approximation problem. Then, based on the Robbins-Monro algorithm, we propose an online algorithm to adaptively learn the optimal sampling policy without channel statistics. Due to the additional transmission randomness brought by the unreliable communication link, we integrate the momentum-based method with the original Robbins-Monro algorithm to reduce the estimation variance of the optimal threshold and improve the convergence rate. 
         \item We prove that the threshold of the proposed algorithm converges to the optimal threshold almost surely. Compared with the previous works \cite{tang2023age, tang2023sampling}, the convergence of the threshold involves the correlated noise from stochastic delays in the adjacent epochs. We prove the almost sure convergence through the ODE method and use the reformulation of the martingale sequence to tackle the correlated noise.
         \item We also provide a theoretical analysis of the convergence rate when there is no frequency constraint and show that the cumulative AoI regret of the online algorithm grows with rate $O(\ln K)$ in Theorem \ref{theo:convergence}.
         \item We verified the optimality of the proposed online algorithm through Le Cam's two-point method. For any online learning algorithm that selects waiting time-based on historical sampling and transmission delay records, the AoI regret under the worst-case delay distribution is lower bounded by $\Omega(\ln K)$. Therefore, the convergence rate of the proposed algorithm is minimax-order optimal.
         \item Finally, simulations are conducted to validate the performance of the proposed online algorithm. The proposed online algorithm consistently achieves lower AoI than the constant waiting policy and converges to the optimal policy under various parameter settings. Through momentum-based variance reduction, we mitigate the impact of the stochastic delay, enhancing the robustness of the proposed algorithm.
     \end{itemize}
    The remaining part of the paper is organized as follows. We describe the system model and formulate the problem in Section \ref{sec:probelm-formulation}. We reformulate the online AoI optimum sampling problem into a stochastic approximation problem in Section \ref{sec:extension-of-the-result} and propose an adaptive learning algorithm. The theoretic analysis of the proposed algorithm is provided in \ref{sec:theo}. Simulation results are presented in Section \ref{sec:simulation}. Finally, the conclusion is drawn in Section \ref{sec:conclusion}.

	\section{Problem Formulation \label{sec:probelm-formulation}}
	\subsection{System Model \label{sec:system-model}}
\begin{figure}[t]
		\centering
		\includegraphics[width=0.9\linewidth]{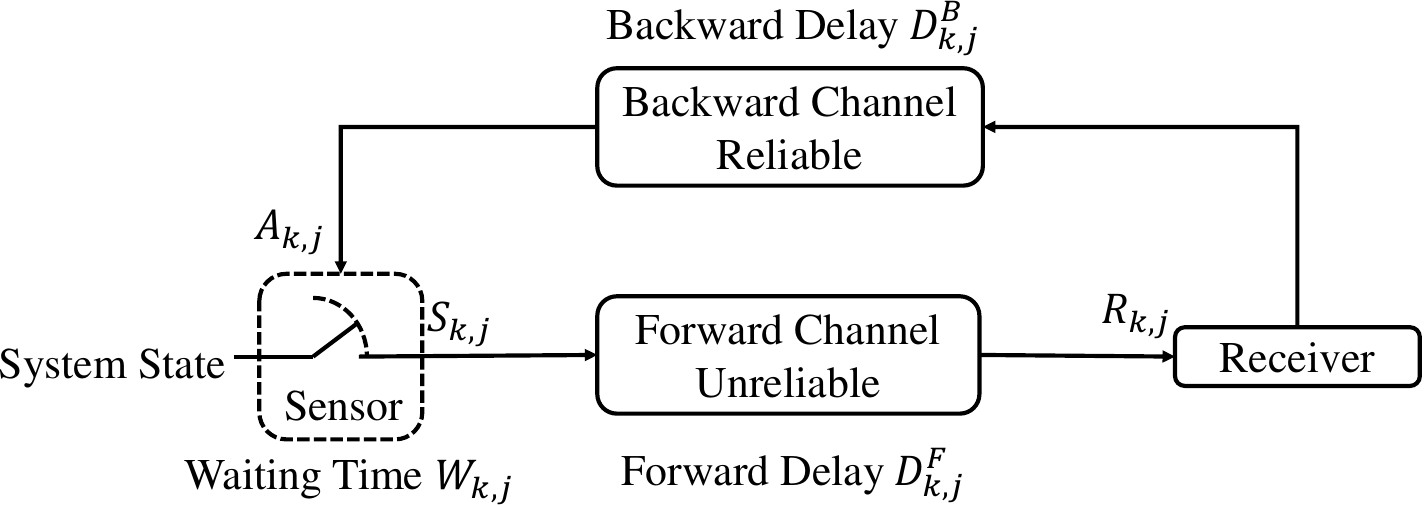}
		\caption{System Model}
		\label{fig:system-model}
	\end{figure}

	We consider a system as demonstrated in Fig.~\ref{fig:system-model}. The system comprises a sensor, a receiver, a forward sensor-to-receiver channel, and a backward receiver-to-sensor channel. The sensor takes a sample of the latest system state and submits the fresh sample to the channel. Due to the fading and interference that exist in the practical environment, we assume that the forward transmission link has a random delay and may suffer from packet-loss. If the transmission succeeds, the receiver immediately sends an acknowledgment (ACK) through the backward channel; otherwise, a negative acknowledgment (NACK) is sent. The feedback transmission channel is error-free and has a random transmission delay.
 
	
	We assume that the packet-loss in the forward transmission is i.i.d with probability $\alpha$. To describe the unreliable transmission easily, we use $k \in \{1, 2, \ldots\}$ to indicate the index of successfully transmitted packets. Then, define the $k$-th epoch to be the time interval between the sampling time of the $k$-th successful transmission and the sampling time of the $(k+1)$-th successful transmission. Due to the packet loss probability, the sensor needs to make $M_k\geq 1$ attempts before the $k+1$-th successful transmission, where $M_k$ follows a geometric distribution with parameter $1-\alpha$. For description of multiple attempts in an epoch, we use the tuple $(k,j)$ to denote the index of $j$-th sampled packet in the $k$-th epoch, where we have $1 \leq j \leq M_k$. Specifically, when $j=1$, the previous sample is successfully delivered to the receiver.
  
	We continue to describe the random delay in both the forward and backward transmission links. Let $S_{k,j}$ be the time-stamp the $(k,j)$ sample is taken. Sample $(k,j)$ experiences a random delay of $D^F_{k,j}$ in the forward channel before reaching the receiver. The reception time is denoted as $R_{k,j}$, at which the receiver attempts to decode the packet and sends an immediate feedback that undergoes a backward random delay $D^B_{k,j}$ that arrives at the sensor at time $A_{k,j}$. We assume that the forward delay $D^F_{k,j}$ and the backward delay $D^B_{k,j}$ are mutually independent and follow their independent and identically distributed probabilities $\mathbb{P}_{\text{FD}}$ and $\mathbb{P}_{\text{BD}}$, respectively. Due to channel propagation delay and time-out constraint, we have the following assumption on the upper and lower bound of the moments of $\mathbb{P}_{\text{FD}}$ and $\mathbb{P}_{\text{BD}}$:
	\begin{assu}
		The probability measures $\mathbb{P}_{\text{FD}}$ and $\mathbb{P}_{\text{BD}}$ are both absolutely continuous on $[\epsilon,\infty)$.\footnote{We assume that each forward and backward transmission has a non-zero link construction time and therefore $D^F\geq\epsilon, D^B\geq \epsilon$. } Moreover, we assume that both the forward and backward transmission delays are fourth-order bounded by a constant $B$, i.e., 
  \begin{equation}
      \mathbb{E}_{\mathbb{P}_{\text{FD}}}[({D^F})^4]\leq B<\infty, \mathbb{E}_{\mathbb{P}_{\text{BD}}}[({D^B})^4]\leq B<\infty.
  \end{equation}
		\label{assu:delay-bound}
	\end{assu}
\begin{remark}~\label{remark:delay-bound}
    Assumption~\ref{assu:delay-bound} implies that the first and second order moment of the forward $D^F$ and backward delay $D^B$ are bounded, i.e., 
    \begin{subequations}
        \begin{align}
        \epsilon^2\leq\mathbb{E}_{\mathbb{P}_{\text{FD}}}[(D^F)^2]\leq\sqrt{\mathbb{E}_{\mathbb{P}}[(D^F)^4]}=\sqrt{B}, \\
        \epsilon\leq\mathbb{E}_{\mathbb{P}_{\text{FD}}}[D^F]\leq\sqrt{\mathbb{E}_{\mathbb{P}}[(D^F)^2]}=B^{1/4}. 
        \end{align}
    \end{subequations}
\end{remark}
    
    Notice that to keep the data fresh, there is no need to submit a new sample if the ACK or NACK of the previous sample has not yet been received \cite{sun2017update}. Therefore, after sampling $(k,j)$-th packet, finding the optimal sampling time for the next packet is equivalent to designing the optimal waiting time $W_{k,j}$ to take a sample after the feedback of the $(k,j)$-th sample is received. Based on the reasoning, following the arrival of the ACK or NACK, the sensor waits for a time period $W_{k,j}$ before acquiring the next sample, where the waiting time $W_{k, j}$ is given by:
    \begin{align}
     W_{k, j} = \left \{
     \begin{aligned}
         &S_{k,j+1}-A_{k, j}, j<  M_{k-1};\\
         &S_{k+1, 1}-A_{k-1, M_{k-1}}, j=M_{k-1}.
     \end{aligned} \right.  
    \end{align}
    The duration of waiting time $W_{k,j}$ is decided by our sampling policy and is assumed to be bounded.

	\subsection{Age of Information}
	We measure how fresh the data is at the receiver via the metric Age of Information (AoI). According to the definition \cite{yates2021age}, AoI is the time difference between the current time and the generation time of the freshest sample. Note that only the first delivered packet in an epoch is successfully transmitted. Then, the AoI $A(t)$ of the current time $t$ is defined as 
	\begin{equation}
		A(t) \triangleq t-\max_k\{S_{k,1}:R_{k,1}\leq t\}.
	\end{equation}
	
\begin{figure}[t]
		\centering
		\includegraphics[width=0.9\linewidth]{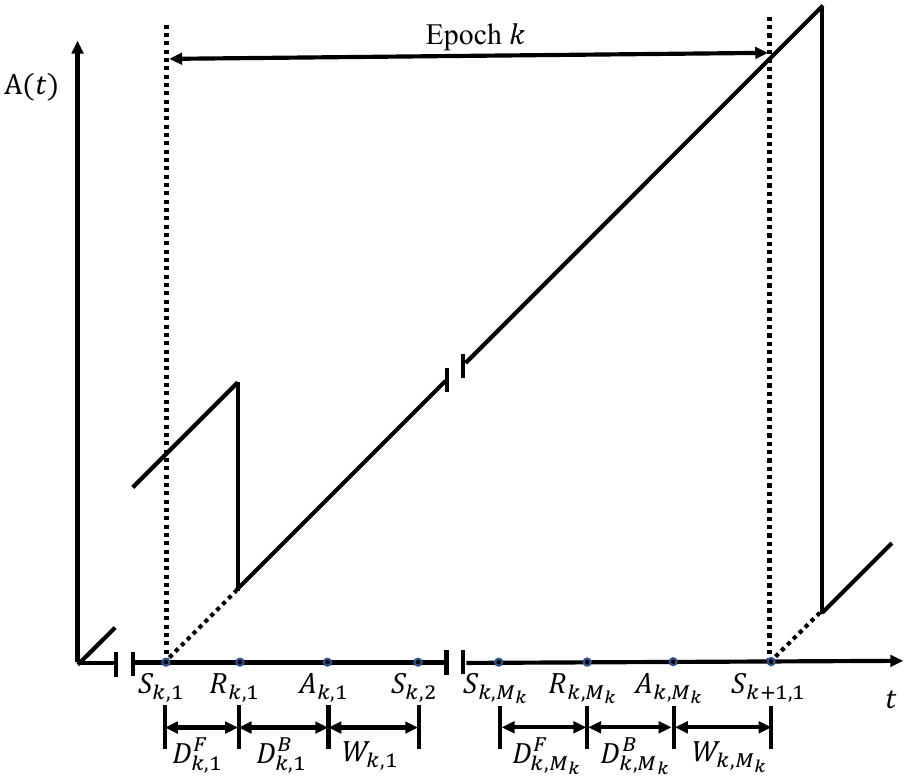}
		\caption{AoI Evolution}
		\label{fig:aoi-evolution}
	\end{figure}

	A sample path of AoI evolution is depicted in Fig.~\ref{fig:aoi-evolution}. After a successful delivery, the AoI decreases to the transmission delay $D^F_{k+1,1}$ of the first sample at the $(k+1)$-th epoch. Otherwise, AoI grows linearly. 
	
	\subsection{Problem Formulation}
	Our objective in this work is to design a sampling policy $\pi$ that selects the waiting time before each transmission to minimize the average AoI when the delay distributions $\mathbb{P}_{\text{FD}}$ and $\mathbb{P}_{\text{BD}}$ and the packet loss probability $\alpha$ are unknown. 
    We only consider the causal policies $\Pi$ in which policy $\pi$ selects a series of waiting time $\{W_{k,j}\}$ based on the history information, i.e., the previous forward and backward delays and the transmission status. Due to the energy constraint, we require that the sampling frequency should be under a certain threshold. Therefore, the optimization problem is formulated as follows:
	\begin{pb}
		\label{pb:problem-1}
		\begin{equation}
			\begin{aligned}
				\text{AoI}_{\text {opt }}=&\inf  _{\pi \in \Pi} \limsup _{T \rightarrow \infty} \frac{1}{T} \mathbb{E}\left[\int_0^T A(t) \text{d}t\right], \\
				&\text {s.t.} \limsup_{T \rightarrow \infty} \frac{1}{T} \mathbb{E}[C(T)] \leq f_{\text{max}} 
			\end{aligned}
		\end{equation}
	\end{pb}
	where $C(T)$ is the total number of samples taken in $[0,T]$.

	\section{Problem Resolution\label{sec:extension-of-the-result}}
 Finding the online sampling algorithm that resolves Problem \ref{pb:problem-1} is divided into two steps: In Section~\ref{sec:sa}, assuming that the transmission statistics $\mathbb{P}_{\text{FD}}, \mathbb{P}_{\text{BD}}, \alpha$ is known, we will review the threshold structure of the AoI minimum sampling policy, and formulate the search for the optimum threshold into a stochastic approximation problem. Next, we utilize the Robbins-Monro algorithm to solve the stochastic approximation problem and improve the stability of the algorithm using a momentum method \ref{sec:alg-arr}.

 \subsection{A Stochastic Approximation Perspective}\label{sec:sa}
 We will first review the properties of the optimum sampling policy in \cite{pan2023optimal}. 
	\begin{corollary}\cite[Theorem 1 Restated]{pan2023optimal}
		A policy $\pi = \{W_{k,j}\}$ is \textit{stationary} and \textit{deterministic} if each waiting time is selected by $W_{k,j}=w(D_{k, j}^F, D_{k, j}^B)$, where $w:\mathbb{R}\times\mathbb{R}\mapsto\{\mathbb{R}^+\cup 0\}$ is a deterministic function from the previous forward and backward transmission delay\footnote{For example, $W_{2,1}$ is a function of $D^B_{2,1}$ and $D^F_{2,1}$ only.}. Moreover, there exists a stationary deterministic policy that solves Problem~\ref{pb:problem-1}, where the waiting time $W_{k,j}$ is selected by:
  \begin{align}
      W_{k,j}=\begin{cases}
          w(D_{k, 1}^F+D_{k, 1}^B), j=1;\\
          0, j > 1. 
      \end{cases}
      \label{eq:optimal-policy-general-case}
  \end{align}
		\label{crl:opt-pol}
	\end{corollary}
 
 Next, we will reformulate the search for the optimum waiting time selection function $w$ into a stochastic approximation problem. For any sampling that satisfies \eqref{eq:optimal-policy-general-case}, the ACK/NACK of sample $(k, 1)$ will be received by $D_k^{\text{a}}:=D_{k, 1}^F+D_{k, 1}^B$ after it is sampled, and then there will be a total of delay $D_k^{\text{v}}:=\sum_{j=2}^{M_k}(D_{k,j}^F+D_{k,j}^B)$ before the end of epoch $k$. Here $D_k^{\text{a}}$ and $D_k^{\text{v}}$ can be viewed as the \emph{actual} and additional ``\emph{virtual}'' delay of sample $(k,1)$. Moreover, $D_k^{\text{a}}$ and $D_k^{\text{v}}$ are i.i.d in each epoch $k$ because both the number of transmission times $M_k$ and the forward, backward transmission delay $D_{k,j}^F, D_{k,j}^B$ are i.i.d. With the introduce of the additional virtual delay, we can turn the time-average AoI minimization problem into the following fractional programming problem, which can be solved by the classical Dinkelbach's Transform \cite{dinkelbach1967nonlinear}. 
\begin{pb}[Fractional Programming Reformulation]
		\label{pb:problem-3}
		\begin{subequations}
			\begin{align}
				\text{AoI}_{\text{opt}}=&\inf_{w:\mathbb{R}\mapsto\{\mathbb{R}^+\cup 0\}}\Big(\mathbb{E}[D^F]+\mathbb{E}[D^{\rm v}]\nonumber\\
       \hspace{1cm}&+\frac{\frac{1}{2}\mathbb{E}\left[(D^{\text{a}}\!+\!w(D^{\text{a}}))^2\right]\!+\!\frac{1}{2}\mathbb{E}[{D^{\text{v}}}^2]\!-\!\mathbb{E}[D^{\text{v}}]^2}{\mathbb{E}[D^{\text{a}}+w(D^{\rm a})]+\mathbb{E}[D^{\text{v}}]}\Big),\\
				&\text { s.t. } \mathbb{E}[D^{\rm a}+w(D^{\rm a})]+\mathbb{E}[D^{\rm v}] \geq \frac{\mathbb{E}\left[M\right]}{f_{\text{max} }}. \label{eq:freqcons}
			\end{align}	
		\end{subequations}
	\end{pb}
     The detailed derivation is in Appendix \ref{sec:reformulation} of the supplementary material. Notice that for any stationary deterministic policy $\pi$ that satisfies the sampling frequency constraint \eqref{eq:freqcons}, the expected AoI achieved by $\pi$ is larger than $\text{AoI}_{\text{opt}}$, i.e.,
     	\begin{align}
		&\mathbb{E}[D^F]\!+\!\mathbb{E}[D^{\rm v}]\!+\!\frac{\frac{1}{2}\mathbb{E}\left[\left(D^{\rm a}+w(D^{\rm a})\right)^2\right]+\frac{1}{2}\mathbb{E}[{D^{\rm v}}^2]-\mathbb{E}[D^{\rm v}]^2}{\mathbb{E}[D^{\rm a}+w(D^{\rm a})]+\mathbb{E}[D^{\rm v}]}\nonumber\\
       &\geq \text{AoI}_{\text{opt}}.
		\label{eq:delta_pi_star-NB}
	\end{align}

 Then deducting $(\mathbb{E}[D^F]+\mathbb{E}[D^{\rm v}])$ and multiplying $\mathbb{E}[D^{\rm a}+w]+\mathbb{E}[D^{\rm v}]$ on both sides of \eqref{eq:delta_pi_star-NB}, Dinkelbach's transform \cite{dinkelbach1967nonlinear} enables us to find the optimum waiting time selection function $w$ of Problem~\ref{pb:problem-3} by solving the following convex optimization problem:
 	\begin{pb}[Convex Optimization]
		\label{pb:problem-4}
		\begin{subequations}
			\begin{align}
				\rho^\star:=&\min\Big(\frac{1}{2}\mathbb{E}\left[ \left(D^{\rm a}+w(D^{\rm a})\right)^2\right]\nonumber\\
    &\hspace{1cm}-(\text{AoI}_{\text{opt}}-\mathbb{E}[D^F]-\mathbb{E}[D^{\rm v}])\mathbb{E}[D^{\rm a}+w(D^{\rm a})]\nonumber\\
    &\hspace{1cm}+\frac{1}{2}\mathbb{E}[{D^{\rm v}}^2]-(\text{AoI}_{\text{opt}}-\mathbb{E}[D^F])\mathbb{E}[D^{\rm v}]\Big),\nonumber\\
				\text { s.t. } &\mathbb{E}\left[D^{\rm a}+w(D^{\rm a})+D^{\rm v}\right] \geq \frac{\mathbb{E}\left[M\right]}{f_{\text{max} }}.\label{eq:freqcons-2}
			\end{align}	
		\end{subequations}
	\end{pb}

 Moreover, the waiting time selection policy $w$ is optimum if and only if $\rho^\star=0$. 
 
 To solve the constrained optimization problem~\ref{pb:problem-4}, we utilize the Lagrange method to obtain the optimal policy under the frequency constraint with dual optimizers $\nu \geq 0$. The Lagrange function is as follows:
	\begin{align}
		\mathcal{L}(\nu, w):=& \frac{1}{2}\mathbb{E}\left[\left(D^{\rm a}+w(D^{\rm a})\right)^2\right]\notag\\
		&-(\text{AoI}_{\text{opt}}-\mathbb{E}[D^F]-\mathbb{E}[D'])\mathbb{E}[D^{\rm a}+w(D^{\rm a})]\nonumber\\
  &+\frac{1}{2}\mathbb{E}[{D^{\rm v}}^2]-(\text{AoI}_{\text{opt}}-\mathbb{E}[D^{\rm v}])\mathbb{E}[D^{\rm v}]\nonumber\\
		&+\nu\left(\frac{\mathbb{E}[M]}{f_{\text{max} }}-\mathbb{E}[D^{\rm a}+w(D^{\rm a})]-\mathbb{E}[D^{\rm v}]\right).
		\label{eq:lagrange-NB-beta}
	\end{align}

 Through the KKT condition, the optimum policy $w^\star$ should be selected to minimize function $\mathcal{L}(\nu, w)$ and the optimum value should satisfy $\mathcal{L}(\nu^\star, w)=0$ according to Dinkelbach's transform. 
	\begin{proposition}
		The optimal function $w^{\star}$ denoted in \eqref{eq:optimal-policy-general-case} is as follows
		\begin{align}
			w^{\star}(D^{\rm a}_k)=
   \left(\text{AoI}_{\text{opt}}-\mathbb{E}[D^F]-\mathbb{E}[D^{\rm v}]+\nu-D^{\rm a}_k\right)^{+}.
			\label{eq:optimum-policy}
		\end{align}\label{prop:opt-pi}
  \end{proposition}

  	The detailed proof is provided in Appendix \ref{sec:prop-opt-pi} of the supplementary material. For simplicity, denote $\gamma^\star:=\text{AoI}_{\text{opt}}-\mathbb{E}[D^F]-\mathbb{E}[D^{\rm v}]$. The sampling policy provided in Corollary~\ref{prop:opt-pi} has a threshold structure in the sense that if the summation of the forward and backward transmission delay is larger than threshold $\gamma^\star+\nu^{\star}$, the sensor will take a new sample immediately; otherwise, the sensor will wait for $\gamma^\star+\nu^{\star}-D_k^{{\rm a}}$ before taking a new sample. The waiting time selection function is:
   \begin{equation}
       w^\star(D_k^{\rm  a})=(\gamma^\star+\nu^\star-D_k^{\rm a})^+. 
   \end{equation}

   It then remains to find the optimum parameter $\gamma^\star$ that minimizes the Lagrange function \eqref{eq:lagrange-NB-beta} when under $\nu^\star$. By Dinkelbach's transform, we know that under the optimum policy, $\mathcal{L}(\nu^\star, w^\star)=0$. Therefore, we have:
	\begin{align}
		0=&\mathcal{L}(\nu^\star, w^\star)\nonumber\\
  =& \frac{1}{2}\mathbb{E}\left[\left(D^{\rm a}+(\gamma^\star+\nu^\star-D^{\rm a})^+\right)^2\right]\notag\\
		&-(\text{AoI}_{\text{opt}}-\mathbb{E}[D^F]-\mathbb{E}[D^{\rm v}])\mathbb{E}[D^{\rm a}+(\gamma^\star+\nu^\star-D^{\rm a})^+]\nonumber\\
  &+\frac{1}{2}\mathbb{E}[(D^{\rm v})^2]-(\text{AoI}_{\text{opt}}-\mathbb{E}[D^F])\mathbb{E}[D^{\rm v}]\nonumber\\
		&+\nu\left(\frac{\mathbb{E}[M]}{f_{\text{max} }}-\mathbb{E}[D^{\rm a}+(\gamma^\star+\nu^\star-D^{\rm a})^+]-\mathbb{E}[D^{\rm v}]\right)\nonumber\\
  \overset{(a)}{=}&\frac{1}{2}\mathbb{E}\left[\max\{D^{\rm a}, \gamma^\star+\nu^\star\}^2\right]-\gamma^\star\left(\mathbb{E}[\max\{D^{\rm a}, \gamma^\star+\nu^\star\}]\right.\nonumber\\
  &\left.+\mathbb{E}[D^{\rm v}]\right)+\underbrace{\frac{1}{2}\mathbb{E}[(D^{\rm v})^2]-\mathbb{E}[D^{\rm v}]^2}_{=:N},\label{eq:opt-sa}
	\end{align}
 where equality $(a)$ is obtained by the KKT condition $\nu^\star\left(\frac{\mathbb{E}[M]}{f_{\text{max} }}-\mathbb{E}[D^{\rm a}+w^\star(D^{\rm a})]-\mathbb{E}[D^{\rm v}]\right)=0$ and  $N:=\frac{1}{2}\mathbb{E}[{D^{\rm v}}^2]-\mathbb{E}[D^{\rm v}]^2$ is a constant.

Equality \eqref{eq:opt-sa} enables us to reformulate a stochastic approximation problem for finding the optimum threshold $\gamma^\star$ and $\nu^\star$. Let function $g_{\nu}(\gamma; D^{\rm a}, D^{\rm v})$ be:
   		\begin{align}
			g_{\nu}(\gamma; D^{\rm a}, D^{\rm v})\nonumber=&\frac{1}{2}\max\{D^{\rm a}, \gamma+\nu\}^2\nonumber\\
   &-\gamma\left(\max\{D^{\rm a}, \gamma+\nu\}+D^{\rm v}\right).
			\label{eq:compute-g}
		\end{align}
   
When the dual optimizer is taken at $\nu^\star$, finding the optimum threshold  $\gamma^\star$ is equivalent to finding the root of the following equation
\begin{equation}
    \overline{g}_{\nu^\star}(\gamma):=\mathbb{E}_{D^{\rm a}, D^{\rm v}}[g_{\nu^\star}(\gamma; D^{\rm a}, D^{\rm v})]=-N. 
    \label{eq:optimal-gamma-new}
\end{equation}

As is proved in \cite[Lemma 1]{pan2023optimal}, the function $\bar{g}(\gamma)$ is monotonic decreasing and concave. 
	
To facilitate the search of  $\gamma^\star$ when the delay distribution $D^F$ and $D^B$ are unknown, we will compute the upper and lower bound of $\gamma^\star$ using the upper and lower bound of the expectation on $D^F, D^B$ derived from Assumption~\ref{assu:delay-bound} and Remark~\ref{remark:delay-bound}. The result is provided in Lemma \ref{lemma:gamma-bound}. 
 
	\begin{lemma}
		The optimal $\gamma^{\star}$ can be bounded by $\gamma_{\text{lb}}\leq \gamma^{\star} \leq \gamma_{\text{ub}}$, where 
		\begin{align*}
			\gamma_{\text{lb}} &:= \max\{\frac{1}{2}(\overline{D^F}_{\text{lb}}+\overline{D^B}_{\text{lb}}-\overline{D^{\rm v}}_{\text{ub}}), 0\}, \\
			\gamma_{\text{ub}} &:= \frac{\frac{1}{2}\overline{H}_{\text{ub}}+\overline{D}_{\text{ub}}\frac{1}{f_{\text{max}}} + \frac{1}{f_{\text{max}}^2}}{\overline{D}_{\text{lb}}+\frac{1}{f_{\text{max}}}}-\overline{D^{\rm v}}_{\text{lb}},
		\end{align*}
        where 
        \begin{subequations}
            \begin{align}
            &\overline{D}_{\text{lb}}\!=\!\overline{D^F}_{\text{lb}}\!+\!\overline{D^B}_{\text{lb}}\!+\!\overline{D^{\rm v}}_{\text{lb}},\label{eq:bound-delay-1}\\
                &\overline{D}_{\text{ub}}\!=\!\overline{D^F}_{\text{ub}}\!+\!\overline{D^B}_{\text{ub}}\!+\!\overline{D^{\rm v}}_{\text{ub}}\label{eq:bound-delay-2}, \\
                &\overline{H}_{\text{ub}}=\overline{(D^F\!+\!D^B\!+\!D^{\rm v})^2}_{\text{ub}}.\label{eq:bound-delay-3}
            \end{align}
        \end{subequations}
		\label{lemma:gamma-bound}
	\end{lemma}
    The proof for Lemma \ref{lemma:gamma-bound} is in Appendix \ref{appd:proof-gamma-bound} of the supplementary material.

 \subsection{An Online Algorithm}\label{sec:alg-arr}
    
	When the channel statistics are unknown, we can estimate the optimal threshold $\gamma^{\star}+\nu^{\star}$ by the Robbins-Monro method. Notice that there is a constant term $N$ defined in \eqref{eq:opt-sa}, which is a combination of the mean and second order moment of $D^{{\rm v}}$. We use $\gamma_k$, $\mu_k$ and $m_k$ to denote our guess about $\gamma^\star$, $\mathbb{E}[D^{\rm v}]$ and $\mathbb{E}[{D^{\rm v}}^2]$ in epoch $k$, which are initialized by $\gamma_0\in \text{Uni}([\gamma_{\text{lb}}, \gamma_{\text{ub}}])$, $\mu_0 = m_0=0$. Notice that $\nu^\star$ is the dual optimizer that guarantees the sampling frequency should be satisfied. Therefore, we approximate $\nu$ by maintaining a sequence $U_k$ to record the sampling frequency debt up to epoch $k$ similar to the Drift-Plus-Penalty framework \cite{neely2010stochastic}. The algorithm operates as follows:

    \begin{itemize}
		\item \textbf{Step 1: Determine the start of epoch $k$:} If the feedback the sensor received from the receiver is an ACK, it means that the newly sampled status information has been received successfully by the receiver. Receiving the $k$-th ACK indicates that the epoch $(k-1)$ is finished, and we have completed the first transmission of epoch $k$ successfully. The sampler computes the virtual delay $D_{k-1}^{\rm v}=\sum_{j=2}^{M_k}(D_{k-1, j}^F+D_{k-1, j}^B)$ of the $(k-1)$-th received sample and actual delay $D_k^{\rm a}=D_{k, 1}^F+D_{k, 1}^B$ of the $k$-th received sample. Then, we update the estimation $\mu_k$ and $m_k$ as follows:
  \begin{subequations}
      \begin{align}
          \mu_k=&\mu_{k-1}+\frac{1}{k}(D_{k}^{\rm v}-\mu_{k-1}),\label{eq:update-mu-k}\\
          m_k=&m_{k-1}+\frac{1}{k}({D_k^{\rm v}}^2-m_{k-1}). \label{eq:update-m-k}
      \end{align}
      Since epoch $k-1$ is finished, we can update the sampling frequency debt up to the beginning of epoch $k$ by:	
      \begin{align}U_{k}\!=\!\Bigg(U_{k-1}\!+\!\bigg[\frac{M_{k-1}}{f_{\text{max}}}\!
			-\!(D_{k-1}^{\rm a}+\!W_{k-1,1}\!+\!D^{\rm v}_{k-1})\bigg]\bigg)^{+}.
			\label{eq:update-U-k}
		\end{align}
		Then, we set $\nu_{k} = \frac{1}{V}U_{k}$ as the dual variable.
  \end{subequations}

        \item \textbf{Step 2: Update $\gamma_{k+1}$ using the Robbins-Monro algorithm:} Assuming that our current estimation about the constant $N_k=\frac{1}{2}m_k-\mu_k^2$ and the dual optimizer $\nu_k$ are accurate, we then proceed to find the root of equation \eqref{eq:opt-sa} by the Robbins-Monro algorithm. Recall that $\gamma_{\rm lb}$ and $\gamma_{\rm up}$ are the upper and lower bound of the target parameter $\gamma$, to find the root of function $\mathbb{E}[g_{\nu}(\gamma;D^{\rm a}, D^{\rm v})]+N=0$ when function $\overline{g}_{\nu}(\cdot; D^{\rm a}, D^{\rm v})$ and is concave and monotonic decreasing, whenever a realization $D^{\rm }_k, D^{\rm v}$ arrives, Kusher \emph{et al.} \cite{kushner1997stochastic} propose to update the target parameter $\gamma_k$ by:
        \begin{align}
            \gamma_k=&\left[\gamma_{k-1}+\eta_k\left(g_{\nu_k}(\gamma_{k-1}; D_k^{\rm a}, D_k^{\rm v})+N_k\right)\right]_{\gamma_{\rm lb}}^{\gamma_{\rm ub}}\nonumber\\
            \overset{(a)}{=}&\Big[\gamma_{k-1}+\eta_k\Big(\frac{1}{2}\max\{D_k^{\rm a}, \gamma_k+\nu_k\}^2-\gamma_k\nonumber\\
            &\cdot\left(\max\{D_k^{\rm a}, \gamma_k+\nu_k\}+D_k^{\rm v}\right)+\frac{1}{2}m_k-\mu_k^2\Big)\Big]_{\gamma_{\rm lb}}^{\gamma_{\rm ub}},\label{eq:update-gamma}
        \end{align}
        where equality $(a)$ is obtained by definition of function $g_{\nu}(\gamma; D^{\rm a}, D^{\rm v})$ from \eqref{eq:compute-g}, $[\cdot]_a^b=\min\{b,\max\{\cdot,a\}\}$ and $\{\eta_k\}$ is a set of convergence sequences selected to be:
		\begin{equation}
			\eta_k=\begin{cases}\frac{1}{2\overline{D}_{\text{lb}}},&k=1;\\ \frac{1}{(k+2)\overline{D}_{\text{lb}}},&k\geq2.\end{cases}\label{eq:steprule}
		\end{equation}

  		\item \textbf{Step 3: Sampling:} After updating $\gamma_k$ and $\nu_k$, we select waiting time $W_{k,1}$ as stated in Proposition \ref{prop:opt-pi}:
		\begin{equation}
			W_{k,1}=(\gamma_{k}+\nu_{k} -D^{\rm a}_{k})^{+}.
			\label{eq:waiting-time-1}
		\end{equation}
  If the feedback we receive from the receiver is a NACK, the recently sampled packet has been lost and we will take a new sample immediately, i.e., the waiting time is selected as zero, i.e.,
		\begin{equation}
			W_{k,j}=0 \quad j=2,3, \cdots.
			\label{eq:waiting-time-2}
		\end{equation}
  And we record that the number of transmissions in epoch $k$ increases by one.
        When receiving an ACK, it indicates the end of epoch $k+1$, and the algorithm will go back to step 1.
	\end{itemize}

	The proposed algorithm is summarized in Algorithm \ref{alg:online-alg}.

\begin{algorithm}
\caption{Proposed Online Algorithm\label{alg:online-alg}}

\begin{algorithmic}[1]
\raggedright
\STATE \textbf{Input:} Frequency Constraint $f_{\text{max}}$, Time $T$, hyper-parameter $V$.
\STATE \textbf{Initialization: }Set $\gamma_0=0, \mu_0=0, m_0=0$ and $U_0=0$
\STATE \textbf{First Sample:} Take a sample immediately at $t=0$ and send it to the receiver
\WHILE{$t\leq T$}
\IF{ACK is received}
\STATE $k\leftarrow k+1$\hfill\COMMENT{A new epoch begins because of ACK}
\STATE $D_k^{\rm a}\leftarrow D_{k, 1}^F+D_{k, 1}^B$\hfill\COMMENT{Step 1: Compute $D_k^{\rm a}, D_{k-1}^{\rm v}$}
\STATE $D_{k-1}^{\rm v}\leftarrow \sum_{j=2}^{M_{k-1}}(D_{k, j}^F+D_{k, j}^B)$
\STATE Compute $\mu_k, m_k, U_k$ according to \eqref{eq:update-mu-k}-\eqref{eq:update-U-k}, set $\nu_k=\frac{1}{V}U_k$
\STATE Update $\gamma_k$ according to \eqref{eq:update-gamma}
\STATE Compute waiting time $W_{k, 1}$ by \eqref{eq:waiting-time-1}, wait for $W_{k, 1}$ to take the next sample
\ELSE
\STATE Take a new sample immediately and send it to the destination, $M_k\leftarrow M_k+1$ \hfill\COMMENT{Epoch $k$ continues due to NACK}
\ENDIF
\ENDWHILE
\end{algorithmic}
\label{step2}
\end{algorithm}

	\section{Theoretical Analysis \label{sec:theo}}
    To theoretically evaluate the algorithm's performance, we first give the almost sure convergence property of the estimation error of the optimal sampling threshold $\gamma^{\star}$, i.e., $\gamma_{K}-\gamma^{\star}$, and the time-averaged AoI difference, i.e., $\frac{\int_{0}^{S_{k+1}}A(t)\text{d}t}{S_{k+1}}-\text{AoI}_{\text {opt }}$ as epochs evolve. Then, we characterize the convergence rate of the threshold estimation error and the cumulative AoI regret of the proposed online algorithm up to epoch $K$, i.e., $\mathbb{E}\left[\int_{0}^{S_{K+1}}A(t)\mathrm{d}t\right]-\mathbb{E}\left[S_{K+1}\right]\overline{A}_{w_{\mathbb{P}}^{\star}}$. Finally, we provide the converse bound for convergence to verify the optimality of the proposed online algorithm. We assume that the upper bound of the transmission delay $D^F, D^B$ and the maximum transmission times in an epoch are known, i.e., $D^F\leq D^F_{\text{ub}}< \infty$, $D^B\leq D^B_{\text{ub}}< \infty$, $M \leq M_{\text{ub}} < \infty$. The main results are as follows.

    \begin{theorem}
    \begin{subequations}
        By using the proposed online algorithm, the threshold $\{\gamma_k\}$ converges to the optimal threshold $\gamma^{\star}$ with probability 1, i.e.,
        \begin{align}
            \lim_{k\rightarrow \infty}\gamma_k \overset{a.s.}{=}\gamma^{\star}.
            \label{eq:a-s-gamma}
        \end{align}

        Moreover, the average AoI of the proposed online algorithm converges to the minimum $\text{AoI}_{\text {opt }}$ with probability 1, i.e.,
        \begin{align}
            \lim_{k \rightarrow \infty} \frac{\int_{0}^{S_{k+1}}A(t)\text{d}t}{S_{k+1}}\overset{a.s.}{=}\text{AoI}_{\text {opt }}.
            \label{eq:a-s-age}
        \end{align}
        
    \end{subequations}
    \label{theo:almost-sure-convergence}
    \end{theorem}

    The proof for Theorem \ref{theo:almost-sure-convergence} is provided in Section \ref{sec:proof-almost-sure-convergence}.
    
    In Theorem \ref{theo:convergence}, we provide the convergence rate of the proposed algorithm.
 
	\begin{theorem}
		\begin{subequations}
			When there is no sampling frequency constraint, i.e., $f_{\text{max}}=\infty$, the approximation error $\gamma_K-\gamma^{\star}$ up to epoch $K$ of the proposed algorithm is upper bounded by:
			\begin{equation}
				E[\left(\gamma_{K}-\gamma^{\star}\right)^{2}]\leq\frac{2}{K}\frac{L_{\text{ub}}^{4}}{\overline{D}_{\text{lb}}^{2}}=\mathcal{O}(\frac{1}{K}),
				\label{eq:theo-1}
			\end{equation}
			where the upper bound of epoch length $L_{\text{ub}}=\gamma_{\text{ub}}+M_{\text{ub}}(D^F_{\text{ub}} +D^B_{\text{ub}})$ and the lower bound of average delay is given in \eqref{eq:bound-delay-1}.
   
            In addition, the cumulative AoI regret of the online algorithm up to epoch $K$ is upper bounded by:
			\begin{align}
   &\mathbb{E}\left[\int_{0}^{S_{K+1}}A(t)\mathrm{d}t\right]-\mathbb{E}\left[S_{K+1}\right]\text{AoI}_{\text {opt }} \nonumber\\
   \leq & \mathbb{E}\left[\sum_{k=1}^{K}\left(\gamma_k - \gamma^{\star}\right)^2\right]\nonumber\\
    \leq & 2 \frac{L_{\text{ub}}^4}{\overline{D}_{\text{lb}}^2}\times \left(1+\ln K\right)=\mathcal{O}(\ln K).
				\label{eq:theo-2}
			\end{align}
		\end{subequations}
		\label{theo:convergence}
	\end{theorem}

    The proof of Theorem~\ref{theo:convergence} is in Section \ref{sec:proof-convergence} of the supplementary material.
    \begin{remark}
        As is shown in \eqref{eq:theo-1}, the estimation error of $\gamma^{\star}$ diminishes over time, indicating the online algorithm learns the optimal policy adaptively. \eqref{eq:theo-2} demonstrates that the cumulative AoI regret increases at a sub-linear rate. Therefore, the average AoI difference between the online algorithm and the optimal policy decreases to 0 when epoch $K$ is sufficiently large.
    \end{remark}

	Furthermore, to measure whether the derived convergence bound is tight, we will provide the converse bound of the proposed online algorithm. Because the delay distributions are general, obtaining a point-wise lower bound for each kind of delay distribution is challenging. As an alternative, we use the minimax error bound through Le Cam's two-point method \cite{le2012asymptotic} to derive the lower bound for the general delay distribution. 
 
    Denote $w^{\star}_{\mathbb{P}}$ as the AoI optimal sampling function that selects the optimal waiting time under the joint delay distribution $\mathbb{P}=\mathbb{P}_{\text{FD}} \cdot \mathbb{P}_{\text{BD}}\cdot \mathbb{P}_\alpha$, $\gamma^{\star}_{\mathbb{P}}$ as the optimal sampling threshold without frequency constraint and $\overline{A}_{w_{\mathbb{P}}^{\star}}$ as the minimum time-averaged AoI. We define historical information obtained in epoch $k$ as $\mathcal{H}_k\triangleq\{M_k, D_{k,j}^F, D_{k,j}^B, 1\leq j \leq M_k \}$ and the cumulative historical information up to epoch $K$ as $\mathcal{H}^{\otimes K}=\{\mathcal{H}_1, \mathcal{H}_2, \cdots, \mathcal{H}_K\}$. At the end of each epoch $K$, we denote $\hat{\gamma}(\cdot):\mathcal{H}^{\otimes {K}}\rightarrow \mathbb{R}^{+}$ as an estimator of the optimal threshold based on historical information $\mathcal{H}^{\otimes {K}}$. According to Le Cam's two-point method, we have the following inequality:
    \begin{equation}
			\inf_{\hat{\gamma}}\sup_{\mathbb{P}}\mathbb{E}[(\hat{\gamma}(\mathcal{H}_K)-\gamma_{\mathbb{P}}^{\star})^2]\geq(\gamma_1-\gamma_2)^2\cdot\mathbb{P}_1^{\otimes K}\wedge\mathbb{P}_2^{\otimes K},
			\label{eq:lecam-theo}
    \end{equation}
    where $\mathbb{P} \wedge \mathbb{Q} = \int \min \{\text{d}\mathbb{P}, \text{d}\mathbb{Q}\}$ denotes the total variation affinity between distributions $\mathbb{P}$ and $\mathbb{Q}$.
 
    To derive the minimax error bound for the estimation of $\gamma^{\star}$, the core idea is to construct two joint distribution $\mathbb{P}_1$, $\mathbb{P}_2$, whose $l_1$ distance $|\mathbb{P}_1^{\otimes K}-\mathbb{P}_2^{\otimes K}|_1$ can be upper bounded by a constant, but $(\gamma^\star_{\mathbb{P}_1}-\gamma^\star_{\mathbb{P}_2})^2 \geq O(1/K)$ is difficult to distinguish. The derived minimax estimation error bound is stated in Theorem \ref{theo:minimax-opt}. 
 
	\begin{theorem}
    \begin{subequations}
        The minimax error bound for the estimation of threshold $\gamma^{\star}$ is as follows:
		\begin{equation}
			\min_{\hat{\gamma}}\max_{\mathbb{P}} \mathbb{E}\left[\left(\hat{\gamma}(\mathcal{H}_K)-\gamma^{\star}_{\mathbb{P}}\right)^2\right] \geq \Omega(\frac{1}{K}).
			\label{eq:minimax-1}
		\end{equation}

        In addition, the time average AoI using any casual waiting time selection function $w$ has the following lower bound:
		\begin{align}
			\inf_{w}\sup_{\mathbb{P}}\left(\mathbb{E}\left[\int_{0}^{S_{K+1}}A(t)\mathrm{d}t\right]-\mathbb{E}[S_{K+1}]\overline{A}_{w_{\mathbb{P}}^{\star}}\right)
			=\Omega\left(\ln K\right).
			\label{eq:minimax-2}
		\end{align}
        \end{subequations}
		\label{theo:minimax-opt}
	\end{theorem}

	The detailed proof of Theorem \ref{theo:minimax-opt} is provided in Section \ref{appd:converse-bound-1}.
	\begin{remark}
		The convergence rate of $\mathbb{E}\left[\left(\gamma_{K}-\gamma^{\star}\right)^{2}\right]$ and increase rate of cumulative AoI regret stated in Theorem \ref{theo:convergence} match the converse bounds in inequality \eqref{eq:minimax-1} and inequality \eqref{eq:minimax-2}. Therefore, the proposed online algorithm is minimax-order optimal. Any other casual policies cannot achieve a better convergence rate than the proposed online algorithm.
	\end{remark}
      
    \section{Momentum-based Variance Reduce\label{sec:variance-reduce}}
	In this section, we introduce momentum to the proposed online algorithm to reduce the variance and improve performance.

    Notice that, at each epoch $k$, the proposed online algorithm updates the threshold $\gamma_k$ through \eqref{eq:update-gamma}, i.e., $\gamma_{k}=[\gamma_{k-1}+\eta_{k} B_{k}]^{\gamma_{\text{ub}}}_{\gamma_{\text{lb}}}$, 
	where $B_{k}=g_{\nu_k}(\gamma_{k-1};D^{\rm a}, D^{\rm v}) + N_k$ is associated with the delays in the previous epoch. $D^{\rm a}_{k}$ and $D^{\rm v}_{k}$ are i.i.d samples of the delay distributions. Therefore, $B_{k}$ is an instance of  $\bar{g}_{\nu_k}(\gamma_k)+N$ when delays take $D^{\rm a}_{k}$ and $D^{\rm v}_{k}$ with noise from the random delays. Due to the stochasticity from the delay samples, the evolution of $\gamma_k$ suffers from large oscillation, leading to the slow convergence rate and sub-optimality of the average AoI.



    We aim to reduce the variance during the stochastic approximation through the momentum-based method.
    Similar to the variance-reduce methods in SGD, the momentum-based update of $\gamma_k$ is as follows:
	\begin{subequations}
		\begin{align}
			d_{k}&=(1-a)d_{k-1}+aB_{k}\label{eq:momentum-1},\\
			\gamma_{k}&=\gamma_{k-1}+\eta_{k} d_{k}.\label{eq:momentum-2}
		\end{align}
	\label{eq:momentum}
	\end{subequations}
	In \eqref{eq:momentum}, $d_{k}$ denotes the momentum term and will be used to update the $\gamma_{k}$. $a$ is the momentum factor and $B_{k}$ is the stochastic estimation of $\bar{g}_{\nu_k}(\gamma_{k-1}) +N$ in the current epoch. The momentum-based algorithm utilizes the superposition of previous estimations $B_{k'}, k'\!=\!1, \cdots, k$ to deviate the current update direction to the optimal threshold $\gamma^{\star}$. The single sample $B_{k}$ is associated with stochastic delays and will endure sudden fluctuation. Therefore, the superposition mitigates the impacts of the random delay and the oscillation of $\gamma_k$, leading to the robustness and improved performance of the online algorithm.

	\section{Simulations \label{sec:simulation}}
    We conduct simulations to evaluate the performance of the proposed algorithm. First, we analyze the average AoI performance both with and without the frequency constraint, comparing it to two different policies. Following this, we examine the impact of varying values of the frequency violation sensitivity parameter $V$. Finally, we assess the performance of momentum-based variance reduction techniques to demonstrate the benefits of momentum modification.
	\subsection{Simulation Settings}
	In this subsection, we provide simulation settings. The packet loss probability $\alpha$ is set as $\alpha = 0.1$ for all the experiments. We consider that the forward and backward transmission delays follow one of the heavy-tailed distributions, i.e., log-normal distribution parameterized by $\mu$ and $\sigma$, which has the density function:
		\begin{equation}
			p(x):=\dfrac{\mathbb{P}_D(\mathrm{d}x)}{\mathrm{d}x}=\dfrac{1}{x\sigma\sqrt{2\pi}}\exp\left(-\dfrac{(\ln x-\mu)^2}{2\sigma^2}\right).
		\end{equation}

	Since the zero-wait policy may not satisfy the sampling frequency constraint, we compare the proposed online algorithm with the following two policies to select the first waiting time in each epoch:
	\begin{enumerate}
		\item A constant wait policy $w_{\text{const}}$ that selects the waiting time by $W_{k,1}=\max\{\frac{\overline{M}}{f_{\text{max}}}-\overline{D^F}-\overline{D^B}-\overline{D^{\rm v}}, 0\}$.
		\item The optimal policy $W_{k,1}=w^{\star}(D^B_{k,1},D^F_{k,1})=\left(\gamma^{\star}+\nu^{\star}-(D^B_{k,1}+D^F_{k,1})\right)^{+},$ where the optimal threshold $\gamma^{\star}+\nu^{\star}$ is computed by \cite{pan2023optimal}.
	\end{enumerate}

	Due to the stochasticity of the channel delays, we repeat the experiment 20 times for each parameter setting and plot the standard variance of the experiments using transparent color. 
	
	\subsection{Sampling without Frequency Constraint}
\begin{figure}
		\centering
		\includegraphics[width=0.9\linewidth]{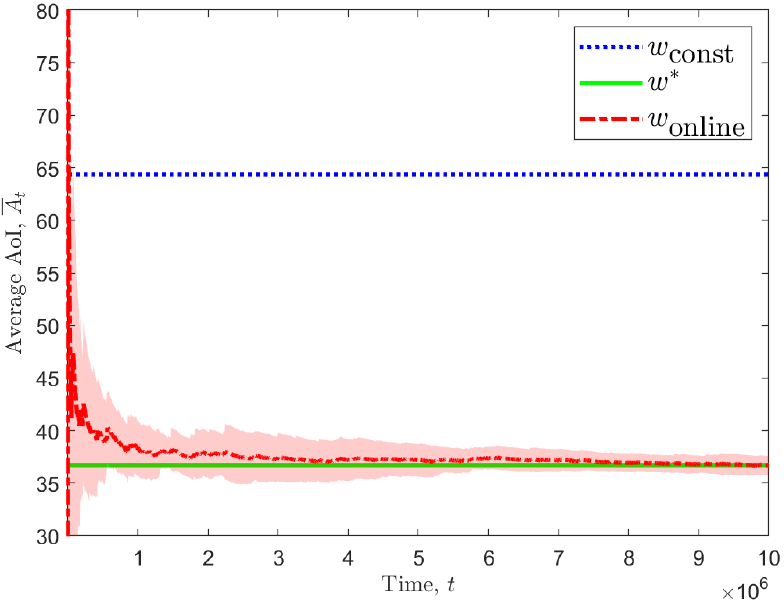}
		\caption{The expected time-average AoI evolution under log-normal(1, 1.8) without frequency constraint}
		\label{fig:sampling-without-freq} 
	\end{figure}

	Fig.~\ref{fig:sampling-without-freq} studies the asymptotic average AoI performance as a function of time using different policies when there is no frequency constraint, i.e., $f_{\text{max}}=\infty$. The parameters of the delays are set to be $\mu_f=\mu_b=1$ and $\sigma_f=1.8, \sigma_b=1$. From Fig.~\ref{fig:sampling-without-freq}, it can be observed that the constant waiting policy has a larger AoI than the proposed online algorithm, which shows the superiority in obtaining data freshness using the proposed online algorithm. In addition, when time $t$ goes to infinity, the average AoI of the online algorithm converges to the minimum AoI obtained by the optimal policy.
	
	\subsection{Sampling with Frequency Constraint}
\begin{figure}
		\centering
		\includegraphics[width=0.93\linewidth]{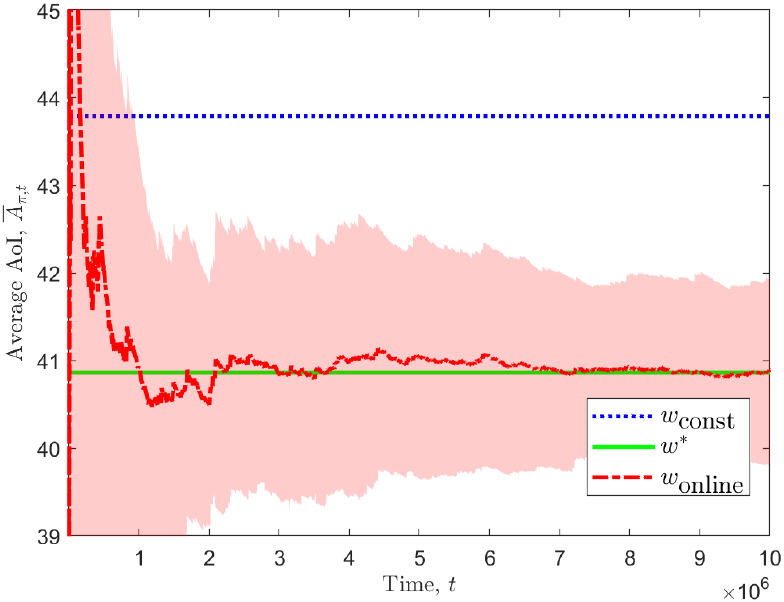}
		\caption{The expected time-average AoI evolution under log-normal(1,1.8) with frequency constraint $f_{\text{max}} = \frac{1}{5(\overline{D^B}+\overline{D^F})}$}
		\label{fig:sim-2}
	\end{figure}

	Fig.~\ref{fig:sim-2} evaluates the asymptotic average AoI performance over time using different policies with frequency constraint, i.e., $f_{\text{max}}=\frac{1}{5(\overline{D^F}+\overline{D^B})}$. The parameters of both the forward and backward delays are set to be $\mu=1$ and $\sigma=1.8$, and the frequency violation sensitivity parameter $V$ is set to be 50. From Fig.~\ref{fig:sim-2}, it can be seen that by using the proposed algorithm, we can achieve a lower AoI performance compared to the constant waiting policy. In addition, similar to the case where there is no frequency constraint, when time $t$ goes to infinity, the average AoI achieved by the proposed online algorithm converges to the minimum AoI. Since the frequency constraint restricts the selection scope of the waiting time, the average AoI gap between the constant-wait policy and the optimal policy becomes smaller than the case without the frequency constraint. 

\begin{figure}
        \centering
        \subfloat[Average sampling interval evolution under log-normal (1, 1.8)]{
			\centering
			\label{fig:sim-V-freq}
			\includegraphics[width=0.93\linewidth]{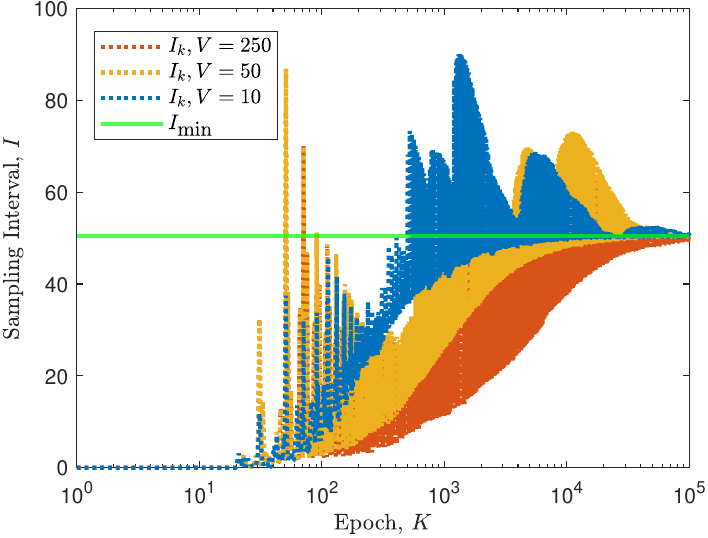}
		}
        \\
        
        \subfloat[Average AoI evolution under log-normal (1, 1.8)]{
			\centering
			\label{fig:sim-V-aoi}
			\includegraphics[width=0.93\linewidth]{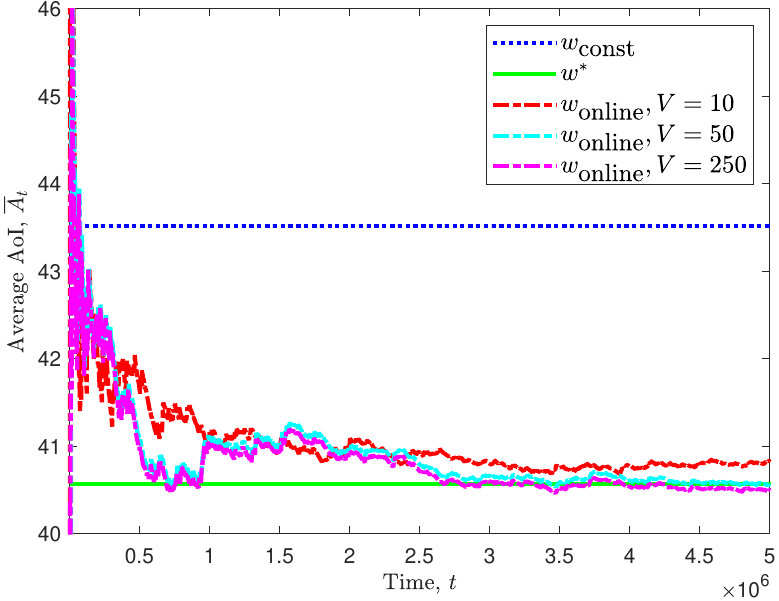}
		}
		\caption{Performance under different value of $V$}
		\label{fig:sim-V}
	\end{figure}

	Fig.~\ref{fig:sim-V} evaluates the evolution of the average AoI and the sampling interval under different values of $V$. In Fig.~\ref{fig:sim-V-freq}, when the number of epochs increases to infinity, the averaged sampling interval with different values of $V$ remains larger or equal to $1/f_{\text{max}}$. Therefore, the frequency constraint of the online algorithm is not violated. In addition, Fig.~\ref{fig:sim-V-aoi} shows that by choosing a larger $V$, the average AoI of the online algorithm converges faster to the optimal average AoI, while by choosing a smaller $V$, the sampling constraint can be satisfied in a shorter time, which is similar to the queueing length-utility trade-off in network utility maximization \cite{neely2008fairness}. The value of $V$ also influences the variance of system performance. With a smaller $V$, we observe a larger turbulence in both AoI evolution and the sampling frequency, which originates from the rapid change in the value of $\nu$. 
	
	\subsection{Momentum-Based Variance Reduction}

\begin{figure}
		\centering
		\includegraphics[width=0.93\linewidth]{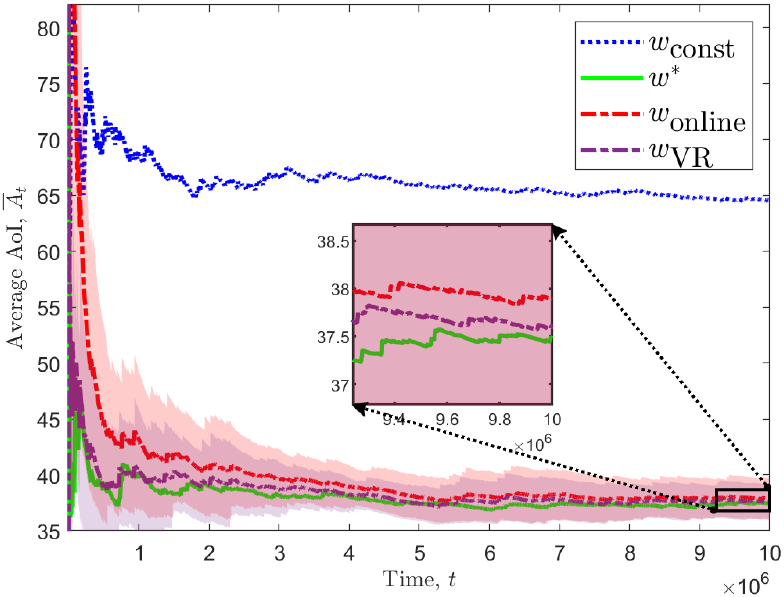}
		\caption{The evolution of average AoI under log-normal(1,1.5)}
		\label{fig:variance-reduce-AoI}
	\end{figure}

	Fig.~\ref{fig:variance-reduce-AoI} displays the evolution of the average AoI as a function of $t$ without frequency constraint, comparing the original online algorithm with the momentum-based algorithm under delay distribution with $\mu=1$ and $\sigma = 1.5$. We adopt the momentum-based method proposed in Section \ref{sec:variance-reduce} with coefficient $a = 0.005$. First, we note that the expected average AoI of the momentum-based algorithm gradually converges to the optimal AoI, exhibiting enhancements over the constant-waiting policy. Moreover, employing the momentum-based variance-reduction technique results in faster convergence of the expected average AoI compared to the original online algorithm, accompanied by a reduction in standard variance.
	

	\begin{figure}
		\centering
		\includegraphics[width=0.93\linewidth]{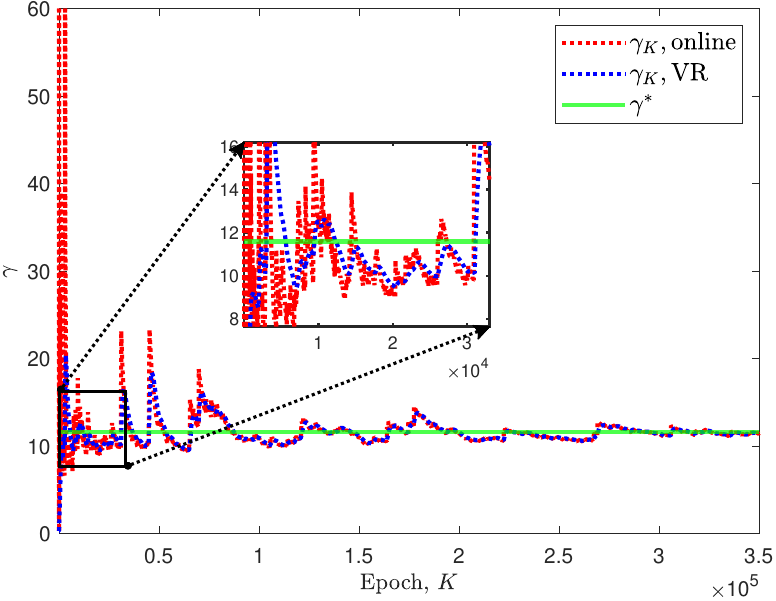}
		\caption{The evolution of $\gamma_k$ under log-normal(1,1.5)}
		\label{fig:variance-reduce-gamma}
	\end{figure}
    Fig.~\ref{fig:variance-reduce-gamma} illustrates the evolution of $\gamma_k$ over the epoch number $k$. We observe that $\gamma_k$ in both the original online algorithm and the momentum-based algorithm converge to the optimal $\gamma^{\star}$ as the epoch number $k$ approaches infinity. However, in the original online algorithm, the evolution of $\gamma_k$ exhibits significant peaks and fluctuations due to stochastic delays. By introducing momentum, we observe that $\gamma_k$ converges to the optimal $\gamma^{\star}$ at a faster rate and with reduced oscillations. This improvement contributes to enhanced AoI performance.

	\section{Proofs of Main Results}
	We provide the proofs for our main results: Theorem 1 (Section \ref{sec:proof-almost-sure-convergence}), Theorem 2 (Section \ref{sec:proof-convergence}) and Theorem 3 (Section \ref{appd:converse-bound-1}). Due to page limitations, the proof of those additional lemmas are provided in the supplementary material \cite{supple}.
	
	\subsection{Proof for Theorem \ref{theo:almost-sure-convergence}\label{sec:proof-almost-sure-convergence}}
	\subsubsection{Proof for \eqref{eq:a-s-gamma}}
	
	 \begin{IEEEproof}
	
		 We study the convergence behavior of sequence $\{\gamma_k\}$ by the ODE method. When there is no sampling constraint, $\nu_k\equiv 0$, the evolution of sequence $\gamma_k$ following \eqref{eq:update-gamma} is as follows:
		 \begin{align}
			 \gamma_k=&\Bigg[\gamma_{k-1}+\eta_k\Bigg(\frac{1}{2}\max\{D_k^{\rm a}, \gamma_k\}^2\nonumber\\
			 &-\gamma_k(\max\{D_k^{\rm a}, \gamma_k\}+D_k^{\rm v})+\frac{1}{2}m_k-\mu_k^2\Bigg)\Bigg]_{\gamma_{\rm lb}}^{\gamma_{\rm ub}}\nonumber\\
			 =&\left[\gamma_{k-1}+\eta_k(g_0(\gamma_k; D_k^{\rm a}, D_k^{\rm v})+\frac{1}{2}m_k^2-\mu_k^2)\right]_{\gamma_{\rm lb}}^{\gamma_{\rm ub}}.\label{eq:pf-gamma-0}
		 \end{align}
		 Define $Z_k$ be the truncating part that forces $\gamma_k$ to interval $[\gamma_{\text{lb}}, \gamma_{\text{ub}}]$, i.e., 
		 \begin{align}Z_k\triangleq&\left[\gamma_{k-1}+\eta_k(g_0(\gamma_k; D_k^{\rm a}, D_k^{\rm v})+\frac{1}{2}m_k^2-\mu_k^2)\right]_{\gamma_{\rm lb}}^{\gamma_{\rm ub}}\nonumber\\
		 &-\left[\gamma_{k-1}+\eta_k(g_0(\gamma_k; D_k^{\rm a}, D_k^{\rm v})+\frac{1}{2}m_k^2-\mu_k^2)\right],
		 \end{align}
		 then the evolution of sequence $\{\gamma_k\}$ in \eqref{eq:pf-gamma-0} be rewritten in the following extended form:
		 \begin{align}
			 \gamma_k=&\gamma_{k-1}+\eta_k\left(\overline{g}_0(\gamma_k)+N\right)\nonumber\\
			 &+\eta_k\left(\underbrace{\frac{k-1}{k}\left(\frac{1}{2}m_{k-1}-\mu_{k-1}^2\right)-\frac{k-1}{k}N}_{=:\beta_k}\right)\nonumber\\
			 &+\eta_k\underbrace{(g_0(\gamma_k;D_k^{\rm a}, D_k^{\rm v})-\overline{g}_0(\gamma_k))}_{=:\delta M_{k, 1}}\nonumber\\
			 &+\eta_k\underbrace{\left(\frac{1}{2}m_k\!-\!\mu_k^2-\left(\frac{k\!-\!1}{k}\left(\frac{1}{2}m_{k-1}\!-\!\mu_{k-1}^2\right)\!-\!\frac{1}{k}N\right)\right)}_{=:\delta M_{k, 2}}\nonumber\\
			 &+\eta_kZ_k.\label{eq:ODEgamma}
		 \end{align}
	
	According to \eqref{eq:ODEgamma}, the update of $\gamma_k$ can be written in the form $\gamma_k=\gamma_{k-1}+\eta_k(Y_k+Z_k)$, where $Y_k$ is defined as follows according to \eqref{eq:ODEgamma}:
	\[Y_{k}:=\overline{g}_0(\gamma_k)+N+\beta_k+\delta M_{k,1}+\delta M_{k,2}. \]
	
	To utilize \cite[p.~95, Theorem 2.1]{kushner1997stochastic}, we will then verify the following conditions for $\{Y_k, \delta M_{k, 1}, \delta M_{k, 2}, \beta_k\}$:
	
	\begin{claim}\label{sec:claim}
		\textbf{(1.1)} $\sup_k \mathbb{E}_k[Y_k^2]< \infty$. 
	
	\textbf{(1.2)} The expectation of $Y_k$ given past observations $\mathcal{H}_{k-1}$ is $\mathbb{E}_k[Y_k]:=\mathbb{E}[Y_k|\mathcal{H}_{k-1}]=\overline{g}_0(\gamma_k)+N+\beta_k$. 
	
	\textbf{(1.3)} Function $\overline{g}_0(\gamma_k)$ is continuous in $\gamma_k$. 
	
	\textbf{(1.4)} The stepsizes $\eta_k$ satisfies $\sum_{k}\eta_k^2<\infty$. 
	
	\textbf{(1.5)} $\sum_k\eta_kb_k<\infty$ with probability 1. 
	
	\end{claim}
	
	Proof of Claim~\ref{sec:claim} is provided in Appendix~G  of the supplementary material \cite{supple}. Therefore,  sequence $\gamma_k$ obtained from \eqref{eq:ODEgamma} will converge to the stationary point of the continuous time ODE:
		 \begin{equation}
			 \dot{\gamma} = \overline{g}_0(\gamma) + N.
			 \label{eq:ODE}
		 \end{equation} 
		 
		 The next step is to show the solution of the ODE in equation \eqref{eq:ODE} converges to $\gamma^{\star}$. Equation \eqref{eq:optimal-gamma-new} implies $\overline{g}_0(\gamma^{\star}) +N= 0$ when $\gamma=\gamma^\star$. Therefore, $\gamma^{\star}$ is an equilibrium point of ODE \eqref{eq:ODE}. To show that the ODE is stationary at $\gamma = \gamma^{\star}$, we use the Lyapunov approach by defining function $V(\gamma):=\frac{1}{2}(\gamma-\gamma^{\star})^2$, whose time derivative $\dot{V}= \frac{d}{dt}V(\gamma(t))$ can be computed by:
		 \begin{equation}
			 \dot{V}=\left(\gamma-\gamma^\star\right)\dot{\gamma}=\left(\gamma-\gamma^\star\right)\left(\overline{g}_0(\gamma)+N\right).
		 \end{equation}
		 
		\begin{lemma}
			For any $\gamma$, the product between distance to the optimal value $\gamma^{\star}$ and the function $\overline{g}_0 + N$ at $\gamma$ is less than 0, i.e.,
			\begin{align}
				 \left(\gamma -\gamma^{\star}\right)\left(\overline{g}_0(\gamma)+N\right)
				\leq  -\left(\gamma -\gamma^{\star}\right)^2\left(\overline{D^{\rm a}} + \overline{D^{\rm v}}\right).
			\end{align}
			\label{lemm:property-lyapunov-function}
		\end{lemma}
		
		 The proof of Lemma \ref{lemm:property-lyapunov-function} is provided in Appendix J of the supplementary material \cite{supple}. According to Lemma \ref{lemm:property-lyapunov-function}, $\dot{V}=\left(\gamma-\gamma^\star\right)\left(\overline{g}_0(\gamma)+N\right)$, the  stability of $\gamma^{\star}$ is verified through Lyapunov theorem.
	
	 \end{IEEEproof}
	
	\subsubsection{Proof for \eqref{eq:a-s-age}}
	
	\begin{IEEEproof}
	
	Notice that the average frame length:
	\begin{align}
		&\liminf_{k\rightarrow \infty} \frac{1}{k}\sum_{k'=1}^k(D^{\rm a}_k +W_{k,1}+D^{\rm v}_k)\nonumber\\
		&\geq \liminf_{k\rightarrow \infty} \frac{1}{k}\sum_{k'=1}^k(D^{\rm a}_k+D^{\rm v}_k)\nonumber\\
		& =\overline{M}(\overline{D^F}+\overline{D^B}) \geq 0, w.p. 1.
	\end{align}
	
	Therefore, to show that the averaged AoI converges to the minimum AoI $\text{AoI}_{\text {opt }}$, it is sufficient to show that $\{\theta_k\}$ converges to 0, where $\theta_k$ is defined as:
		\begin{align}
			\theta_k:=\frac{1}{k}\left(\int_0^{S_{k+1}}A(t)\text{d}t - S_{k+1}\text{AoI}_{\text {opt }}\right).
		\end{align}
		
		The proof of the almost sure convergence of the cumulative age can be divided into two steps. First, we will show that with probability 1, $\{\theta_k\}$ converges to the limit point of an ODE. Second, we will show that 0 is the unique stationary point of the ODE.
		
		To construct the ODE, we will reformulate the evolution of $\{\theta_k\}$ to a recursive form. Recall that $\int_0^{S_{k+1}}A(t)\text{d}t\sum_{k'=1}^k F_{k'}$ and the optimal AoI is expressed as $\gamma^{\star}+\overline{D^{\rm v}}+\overline{D^F}$. We have:
		\begin{align}
			\theta_k=&\frac{1}{k}\left(\int_0^{S_{k+1}}A(t)\text{d}t - \left(\gamma^{\star}+\overline{D^{\rm v}}+\overline{D^F}\right)S_{k+1}\right)\nonumber\\
			=&\frac{1}{k}\sum_{k'=1}^{k}\left(F_{k'} - \left(\gamma^{\star}+\overline{D^{\rm v}}+\overline{D^F}\right)L_k\right)\nonumber\\
			=&\frac{1}{k}\Big(\left(k\!-\!1\right)\theta_{k-1}\!+\!\frac{1}{2}\max\left\{D^{\rm a}_k,\gamma_k\right\}^2+D_k^{\rm v}\max\left\{D_k^{\rm a},\gamma_k\right\}\nonumber\\
			&+\frac{1}{2}(D_k^{\rm v})^2+D_{k,1}^FL_{k-1}- \left(\gamma^{\star}+\overline{D^{\rm v}}+\overline{D^F}\right)L_k\Big) \nonumber\\
			=&\theta_{k-1}\!+\!\frac{1}{k}\Big(\frac{1}{2}\max\left\{D^{\rm a}_k,\gamma_k\right\}^2\!+\!D_k^{\rm v}\max\left\{D_k^{\rm a},\gamma_k\right\}\!-\!\theta_{k-1}\nonumber\\
			&+\frac{1}{2}(D_k^{\rm v})^2+D_{k,1}^FL_{k-1}\!-\! \left(\gamma^{\star}\!+\!\overline{D^{\rm v}}+\overline{D^F}\right)L_k\Big).
		\end{align}
	
		Define $Y_k=\frac{1}{2}\max\left\{D^{\rm a}_k,\gamma_k\right\}^2+D_k^{\rm v}\max\left\{D_k^{\rm a},\gamma_k\right\}+\frac{1}{2}(D_k^{\rm v})^2+D_{k,1}^FL_{k-1}- \left(\gamma^{\star}+\overline{D^{\rm v}}+\overline{D^F}\right)L_k-\theta_{k-1}$. Then, the update of $\theta_k$ can be expressed as:
		\begin{align}
			\theta_k:=\theta_{k-1}+\frac{1}{k}\left(\mathbb{E}\left[Y_k| \mathcal{H}_{k-1}\right]+\left(Y_k-\mathbb{E}\left[Y_k|\mathcal{H}_{k-1}\right]\right)\right).
		\end{align}
	
		Given the historical information $\mathcal{H}_{k-1}$, the conditional expectation of $Y_k$ can be expressed as:
		\begin{align}
			&\mathbb{E}[Y_k\mid\mathcal{H}_{k-1}]\nonumber\\
			=&\mathbb{E}\Big[\frac{1}{2}\max\left\{D^{\rm a}_k,\gamma_k\right\}^2-\gamma_k\left(\max\left\{D_k^{\rm a},\gamma_k\right\}+D_k^{\rm v}\right)-\theta_{k-1}\nonumber\\
			&+D_k^{\rm v}\max\left\{D_k^{\rm a},\gamma_k\right\}+\frac{1}{2}(D_k^{\rm v})^2+D_{k,1}^FL_{k-1}\nonumber\\
			&- \left(\gamma^{\star}+\overline{D^{\rm v}}+\overline{D^F}-\gamma_k\right)L_k\Big]\nonumber\\
			=&\mathbb{E}\Big[\frac{1}{2}\max\left\{D^{\rm a}_k,\gamma_k\right\}^2\!-\!\gamma_k\left(\max\left\{D_k^{\rm a},\gamma_k\right\}\!+\!D_k^{\rm v}\right)\!-\!\theta_{k\!-\!1}|\mathcal{H}_{k-1}\Big]\nonumber\\
			&+\overline{D}^F\left(\mathbb{E}\left[L_{k-1}|\mathcal{H}_{k-1}\right]-l(\gamma_k)\right)+\left(\gamma_k-\gamma^{\star}\right)l(\gamma_k)\nonumber\\
			&+
			\mathbb{E}\Big[D_k^{\rm v}\max\left\{D_k^{\rm a},\gamma_k\right\}+\frac{1}{2}(D_k^{\rm v})^2- \overline{D^{\rm v}}L_k\Big]\nonumber\\
			=&\mathbb{E}\Big[\frac{1}{2}\max\left\{D^{\rm a}_k,\gamma_k\right\}^2-\gamma_k\left(\max\left\{D_k^{\rm a},\gamma_k\right\}+D_k^{\rm v}\right)-\theta_{k-1}\Big]\nonumber\\
			&+\underbrace{\overline{D}^F\!\left(\mathbb{E}\left[L_{k-1}|\mathcal{H}_{k-1}\right]\!-\!l(\gamma_k)\right)}_{:=\beta_{k,1}}\!+\!\underbrace{\left(\gamma_k\!-\!\gamma^{\star}\right)l(\gamma_k)}_{:=\beta_{k,2}}+N.
		\end{align}
	
		Define function 
		\begin{align}
			f(\theta,\gamma;D^{\rm a},D^{\rm v})=&\frac{1}{2}\max\left\{D^{\rm a}_k,\gamma_k\right\}^2\nonumber\\
			&-\gamma\left(\max\left\{D_k^{\rm a},\gamma_k\right\}+D^{\rm v}_k\right)-\theta,
		\end{align}
		and the average over delay $D^{\rm a},D^{\rm v}$ as 
		\begin{align}
			\overline{f}(\theta,\gamma)=\mathbb{E}_{D^{\rm a},D^{\rm v}}\left[f(\theta,\gamma;D^{\rm a},D^{\rm v})\right].
		\end{align}
	
		In the following analysis, we will prove that the sequence $\{\theta_k\}$ converges to the stationary point of an ODE induced by the function $\overline{f}(\theta, \gamma)$. Denote $\delta M_k=Y_k-\mathbb{E}[Y_k\mid \mathcal{H}_{k-1}]$. The recursive update of $\theta$ can be expressed as
		\begin{align}
			\theta_k=\theta_{k-1}\!+\!\frac{1}{k}\left(\overline{f}(\theta_{k-1}, \gamma_k)\!+\!\delta M_k\!+\!\beta_{k,1}\!+\!\beta_{k,2}\!+\!N\right).
			\label{eq:recursive-theta}
		\end{align}
	
		Before proceeding to give the properties of $\{Y_k, \delta M_k,\beta_{k,1},\beta_{k,2}\}$, we will define some variables. Denote $\epsilon_k = \frac{1}{k}$, which can be viewed as the step-size for updating $\theta_k$. Term $\beta_{k,1}$ and $\beta_{k,2}$ can be viewed as two bias terms. Define $t_0=0$ and the cumulative step-size up to epoch $k$ is denoted by $t_k=\sum_{i=0}^{k-1}\epsilon_i$. Therefore, $\ln_k \leq t_k \leq 1+\ln(k-1)$. For $t\geq 0$, let $m(t)$ be the unique value such that $t_{m(t)} \leq t < t_{m(t)+1}$. We have 
		\begin{align}
			m(t)=\lfloor \exp(t)\rfloor.
		\end{align}
		We present the following properties about the recursive equation \eqref{eq:recursive-theta}:
		\begin{claim}
			Sequence $\{Y_k, \delta M_k,\beta_{k,1},\beta_{k,2}\}$ satisfy the following properties:
			
			\noindent \textbf{(2.1)} $\sup_k \mathbb{E}[\mid Y_k\mid ] < \infty$.
			
			\noindent \textbf{(2.2)} $\overline{f}(\theta, \gamma)$ is continuous in $\theta$.
			
			\noindent\textbf{(2.3)} 
			We have the limit for all $\theta$:
			\begin{align}
			   \hspace{-0.5cm} &\lim_{k\to\infty}\Pr\left(\sup_{j\geq k}\max_{0\leq t\leq T}\left|\sum_{i=m(jT)}^{m(jT+t)-1}\epsilon_i\left(f(\theta,\gamma_i)-\overline{f}(\theta)\right)\right|\geq\mu\right)\nonumber\\
				&=0.
			\end{align}
			\noindent \textbf{(2.4)} For each $\mu >0$, we have 
			\begin{align}
				&\lim_{k\rightarrow \infty}\text{Pr}\left(\sup_{j\geq k}\max_{0\leq t\leq T}\left| \sum_{i=m(jT)}^{m(jT+t)-1}\epsilon_i\delta M_i\right|\geq \mu\right)=0.
			\end{align}
			
			\noindent \textbf{(2.5)} The bias sequence satisfies:
			\begin{align}
				\lim_{k\rightarrow \infty}\text{Pr}\left(\sup_{j\geq k}\max_{0 \leq t\leq T}\left| \sum_{i=m(jT)}^{m(jT)+1}\epsilon_i(\beta_{k,1}+\beta_{k,2})\right|\geq \mu\right)
				=0.
			\end{align}
	
			\noindent \textbf{(2.6)} Function $f$ is uniformly bounded for $\theta \in [0, 2L_{\text{ub}}^2]$, $\gamma\in [\gamma_{\text{lb}}, \gamma_{\text{ub}}]$.
			
			\noindent \textbf{(2.7)} For each $\gamma$ we have:
				$|f(\theta_1, \gamma)-f(\theta_2, \gamma)|=|\theta_1-\theta_2|$,
			and $\lim_{|\theta_1-\theta_2| \rightarrow 0}|f(\theta_1, \gamma)-f(\theta_2, \gamma)|=0$.
			
			\noindent \textbf{(2.8)} Sequence $\frac{1}{k}$ satisfies $\sum_{k'=1}^{\infty}\frac{1}{k'}=\infty$.
		\label{sec:claim-2}
		\end{claim}
		
		The proof of claim 2 is provided in Appendix H of the supplementary material \cite{supple}.
		Therefore, according to \cite[p. 140, Theorem 1.1]{yu1997assouad}, with probability 1, sequence $\theta_k$ converges to the limit point of the following ODE:
		\begin{align}
			\dot{\theta}=\overline{f}(\theta, \gamma^{\star})+N.
			\label{eq:ODE-theta}
		\end{align}
	
		Because $\overline{f}(0, \gamma^{\star})+N=0$, and this is the equilibrium point of the ODE in equation \eqref{eq:ODE-theta}. Therefore, $\theta_k$ converges to the equilibrium point with probability 1, and the time-averaged AoI converges to $\text{AoI}_{\text {opt }}$ with probability 1, i.e.,
		\begin{align}
			\lim_{k \rightarrow \infty}\int_0^{S_{k+1}}A(t)\text{d}t - \text{AoI}_{\text {opt }}S_{k+1} \overset{a.s.}{=} 0.
		\end{align}
	
	\end{IEEEproof}

	\subsection{Proof for Theorem \ref{theo:convergence}\label{sec:proof-convergence}}
	
	\subsubsection{Proof for (23a)}
	
	\begin{IEEEproof}
		
		We will use the Lyapunov method. Recall that $V(\gamma)=\frac{1}{2}(\gamma-\gamma^\star)^2$ is the Lyapunov function, we have:
		\begin{align}
			&\mathbb{E}_k\left[V(\gamma_k)\right]-V(\gamma_{k-1})\nonumber\\
			=&\mathbb{E}_k\Big[\left(\left[\gamma_{k-1}+\eta_k (\overline{g}_0(\gamma_{k-1})+N+b_k)\right.\right.\nonumber\\&\left.\left.+\eta_k(\delta M_{k, 1}+\delta M_{k, 2})\right]_{\gamma_{\rm lb}}^{\gamma_{\rm ub}}-\gamma^\star\right)^2\Big]
			-(\gamma_{k-1}-\gamma^\star)^2\nonumber\\
			\overset{(a)}{\leq}&\mathbb{E}_k\Big[\left(\gamma_{k-1}+\eta_k(\overline{g}_0(\gamma_{k-1})+N+b_k)\right.\nonumber\\
			&\left.+\eta_k(\delta M_{k, 1}+\delta M_{k, 2})-\gamma^\star\right)^2\Big]
			-(\gamma_{k-1}-\gamma^\star)^2\nonumber\\
			=&\mathbb{E}_k\left[\left(\gamma_{k-1}+\eta_k(\overline{g}_0(\gamma_{k-1})+N+b_k)-\gamma^\star\right)^2\right]\nonumber\\
			&+2\eta_k\mathbb{E}_k\left[\left(\gamma_{k-1}+\eta_k(\overline{g}_0(\gamma_{k-1})+N+b_k)-\gamma^\star\right)^2\right]\nonumber\\
			&\cdot\mathbb{E}\left[\delta M_{k, 1}+\delta M_{k, 2}\right]
			+\eta_k^2\mathbb{E}\left[\left(\delta M_{k, 1}+\delta M_{k, 2}\right)^2\right]\nonumber
			\\
			&-(\gamma_{k-1}-\gamma^\star)^2\nonumber\\
			\overset{(b)}{\leq }&\mathbb{E}_k\left[\left(\gamma_{k-1}+\eta_k(\overline{g}_0(\gamma_{k-1})+N+b_k)-\gamma^\star\right)^2\right]+\frac{1}{k^2}N_1\nonumber\\
			&-(\gamma_{k-1}-\gamma^\star)^2\nonumber\\
			=&\eta_k\mathbb{E}[\left(\overline{g}_0(\gamma_{k-1})+N\right)(\gamma_{k-1}-\gamma^\star)]+\eta_k\mathbb{E}[b_k](\gamma_{k-1}-\gamma^\star)\nonumber\\
			&+2\eta_k^2\overline{g}_0(\gamma_{k-1})^2+2\eta_k^2\mathbb{E}[b_k^2]+\frac{1}{k^2}N_1\nonumber\\
			\overset{(c)}{\leq }&-\eta_k\left(D^{\rm a}+D^{\rm v}\right)V(\gamma_{k-1})+\eta_k\gamma_{\rm ub}\mathbb{E}[|b_k|]+\eta_k^2 N_2,
			\label{eq:lyapunov-iterate}
		\end{align}
		where $(a)$ is obtained because $\gamma^\star\in[\gamma_{\rm lb}, \gamma_{\rm ub}]$; inequality $(b)$ is obtained because $\delta M_{k, 1}, \delta M_{k, 2}$ is martingale sequences and therefore $\eta_k\mathbb{E}_k((\gamma_{k-1}+\eta_k b_k-\gamma^\star)^2]\mathbb{E}[\delta M_{k, 1}+\delta M_{k, 2}]=0$. As $\gamma_{k-1}$ is upper and lower bounded and the second order of $D_k^{\rm v}$ is upper bounded, $\mathbb{E}[\delta M_{k, 1}^2]$ and $\mathbb{E}[\delta M_{k, 2}^2]$ are all upper bounded. Inequality (c) is from Lemma 2. 
		
		Multiplying inequality \eqref{eq:lyapunov-iterate} from $i=1$ to $k$ yields:
		\begin{align}
			\mathbb{E}\left[V(\gamma_{k+1})\right] \leq& \sum_{i=1}^k\left(\eta_i^2N_2+\eta_k\gamma_{\rm ub}\mathbb{E}[|b_k|]\right)\cdot \prod_{j=i+1}^k(1-\eta_j \overline{D}_{\text{lb}})\nonumber\\
			&+\prod_{i=1}^k(1-\eta_i \overline{D}_{\text{lb}})V(\gamma_0).
		\end{align}
		
		Since the stepsize selected satisfies:
		\begin{align}
			\eta_k \rightarrow 0, \liminf_{k}\min_{n\geq i\geq m(t_k-T)}\frac{\eta_n}{\eta_i}=1,
		\end{align}
		according to \cite[p. 343, Eq. (4.8)]{yu1997assouad}, term $\prod_{i=1}^k(1-\eta_i \overline{D}_{\text{lb}})=\mathcal{O}(\eta_k)$. Therefore,
		\begin{align}
			\sup_k \mathbb{E}\left[\frac{\left(\gamma_k-\gamma^{\star}\right)^2}{\eta_k}\right]=\sup_k\mathbb{E}\left[2V(\gamma_k)/\eta_k\right]=\mathcal{O}(1).
		\end{align}
	
	\end{IEEEproof}

	\subsubsection{Proof for \eqref{eq:theo-2}}
	\begin{IEEEproof}
		
		Firstly, we establish the connection between the accumulation of AoI until epoch $K$ and the threshold different $\left(\gamma_k - \gamma^{\star}\right)^2$, as stated in Lemma \ref{lemma:bound-for-gamma}.
		\begin{lemma}
			According to the definition in \eqref{eq:definition-F}, the cumulative AoI in epoch $k$ can be expressed as 
			$F_k = Q_k+L_{k-1}D^F_{k,1}+\frac{1}{2}(D^{\rm v}_k)^2+D^{\rm v}_k\max\{D^{\rm a}, \gamma_k\}.$ The cumulative AoI until the end of epoch $K$ can be rewritten as a sum of $F_k$:
			$\mathbb{E}\left[\int_{0}^{S_{K+1}}A(t)\text{d}t\right] = \mathbb{E}\left[\sum_{k=1}^{K}F_k\right]$, which
			satisfies the following inequality:
			\begin{equation}
				\!\mathbb{E}\!\left[\sum_{k=1}^K\!\left(F_k \!-\! (\gamma^{\star} \!+\!\overline{D^{\rm v}}\!+\! \overline{D^F})L_k\right)\!\right] \!\leq\! \mathbb{E}\!\left[\sum_{k=1}^{K}\left(\gamma_k \!-\! \gamma^{\star}\right)^2\!\right]\!.
			\end{equation}
			\label{lemma:bound-for-gamma} 
		\end{lemma}
		The proof for Lemma \ref{lemma:bound-for-gamma} is provided in Appendix \ref{appd:proof-gamma-bound}.

		Utilizing Lemma \ref{lemma:bound-for-gamma}, we can upper bound the cumulative AoI regret as follows:
		\begin{align}
			&\mathbb{E}\left[\int_{0}^{S_{K+1}}A(t)\text{d}t - S_{K+1}\text{AoI}_{\text {opt }}\right]\nonumber\\
			=&\mathbb{E}\left[\sum_{k=1}^{K}\left(F_k \!-\! \left(\gamma^{\star} + \overline{D^{\rm v}} + \overline{D^F}\right)L_k\right)\right] 
			\!\leq\! \mathbb{E}\left[\sum_{k=1}^{K}\left(\gamma_k \!-\! \gamma^{\star}\right)^2\right].
			\label{eq:A-k-bound}
		\end{align}
		Next, we will use the upper bound for $\gamma_k$ to derive the cumulative AoI regret bound. Summing up (23a) from $1$ to $K$, we have:
		\begin{align}
			\mathbb{E}\left[\sum_{k=1}^{K}(\gamma_k-\gamma^{\star})^2\right] 
			\leq & 2\frac{L_{\text{ub}}^4}{\overline{D}_{\text{lb}}^2}\left(\sum_{k=1}^{K}\frac{1}{k}\right) \nonumber \\
			\overset{(a)}{\leq} & 2\frac{L_{\text{ub}}^4}{\overline{D}_{\text{lb}}^2}\left(1 + \int_{1}^{K}\frac{1}{k}\text{d}k\right) \nonumber \\
			\overset{(b)}{=} & 2\frac{L_{\text{ub}}^4}{\overline{D}_{\text{lb}}^2}\left(1 + \ln K\right).
			\label{eq:ln-k-bound}
		\end{align}
		where inequality (a) is because $\frac{1}{k} \leq \int_{k^{\prime}=k-1}^{k}\frac{1}{k^{\prime}}\text{d}k^{\prime}$ and equality (b) is the direct result from the integration. Plugging inequality \eqref{eq:ln-k-bound} into \eqref{eq:A-k-bound}, we arrive to the statement of (23b):
		\begin{align}
			&\mathbb{E}\left[\int_{0}^{S_{K+1}}A(t)\text{d}t - S_{K+1}\text{AoI}_{\text {opt }}\right]\nonumber\\
			\leq & \mathbb{E}\left[\sum_{k=1}^{K}(\gamma_k - \gamma^{\star})^2\right]\leq 2\frac{L_{\text{ub}}^4}{\overline{D}_{\text{lb}}^2} \left(1 + \ln K\right),
		\end{align}
		where the last inequality is from $\mathbb{E}\left[\sum_{k=1}^{K}L_k\right] \geq \mathbb{E}\left[\sum_{k=1}^{K} \left(D^{\rm a}_{k}+D^{\rm v}_k\right)\right] \geq K \overline{D}$. 
	
	\end{IEEEproof}
	
	\subsection{Proof of Theorem \ref{theo:minimax-opt}\label{appd:converse-bound-1}}
		\subsubsection{Proof of Equation \eqref{eq:minimax-1}\label{subsec-minimax-1}}
	
				\begin{IEEEproof}
	
			We prove Theorem \ref{theo:minimax-opt} through Le Cam's two-point method. Recall that $\gamma^{\star}$ is the root of \eqref{eq:optimal-gamma-new}, for any joint distribution $\mathbb{P}\triangleq \mathbb{P}_{\text{FD}}\cdot\mathbb{P}_{\text{BD}}\cdot\mathbb{P}_{\alpha}$, the optimal threshold $\gamma_\mathbb{P}^{\star}$ satisfies
			\begin{align}
			\frac{1}{2}\mathbb{E}_{\mathbb{P}}\left[\max\{D^{\rm a}, \gamma_\mathbb{P}^{\star}\}^2\right]
			\!-\!\gamma_\mathbb{P}^{\star}\mathbb{E}_{\mathbb{P}}\left[\max\{D^{\rm a}, \gamma_\mathbb{P}^{\star}\}\!+\!D^{\rm v}\right] \!+\! N = 0.
				\label{eq:optimum-thre}
			\end{align}
			
			Denote $\mathbb{P}_1$ and $\mathbb{P}_2$ as two probability distributions and denote the optimal thresholds for each distribution set as $\gamma_1=\gamma^{\star}_{\mathbb{P}_1}$ and $\gamma_2 = \gamma^{\star}_{\mathbb{P}_2}$, for simplicity. Let $\mathbb{P}_1^{\otimes k}$ and $\mathbb{P}_2^{\otimes k}$ be the distributions of the historical information obtained in epoch $1$ to $k$, i.e., $\mathcal{H}^{\otimes k}=\{\mathcal{H}_1, \cdots, \mathcal{H}_k\}$. Therefore, $\mathbb{P}_1^{\otimes k}$ and $\mathbb{P}_2^{\otimes k}$ are product of distribution of $k$ i.i.d samples drawn from $\mathbb{P}_1$ and $\mathbb{P}_2$, respectively. According to the Le Cam's two-point method \cite{yu1997assouad, le2012asymptotic}, the minimax lower bound of the estimation of $\gamma_k$ satisfies:
			\begin{equation}
				\inf_{\hat{\gamma}}\sup_{\mathbb{P}}\mathbb{E}\left[\left(\hat{\gamma}\left(\mathcal{H}_k\right)-\gamma_{\mathbb{P}}^{\star}\right)^2\right]\geq\left(\gamma_1-\gamma_2\right)^2\cdot\mathbb{P}_1^{\otimes k}\wedge\mathbb{P}_2^{\otimes k},
				\label{eq:lecam-ineq}
			\end{equation}
			where $\mathbb{P} \wedge \mathbb{Q} = \int \min \{\text{d}\mathbb{P}, \text{d}\mathbb{Q}\}$ denotes the total variation affinity between distributions $\mathbb{P}$ and $\mathbb{Q}$.
			
			To obtain the desired lower bound, we want to find two distinct joint distributions of forward, backward transmission delay and packet transmission failure rate $\mathbb{P}_1$ and $\mathbb{P}_2$, such that the difference between the sampling threshold $\left(\gamma_1 - \gamma_2\right)^2$ is large while the total variation affinity $\mathbb{P}_1^{\otimes k}\wedge\mathbb{P}_2^{\otimes k}$ can be lower bounded. 
			
			We consider two distribution sets that share the same delay distributions but differ in the rate of failure. Define $\mathbb{P}_1:=\mathbb{P}_{\text{FD}} \cdot  \mathbb{P}_{\text{BD}}\cdot \mathbb{P}_{\alpha_1}$ and $\mathcal{P}_2:=\mathbb{P}_{\text{FD}} \cdot  \mathbb{P}_{\text{BD}}\cdot \mathbb{P}_{\alpha_2}$, where $\alpha_1$ is greater than $\alpha_2$. In addition, we set $\mathbb{P}_{\text{FD}}=\mathbb{P}_{\text{BD}}=\text{Uni}[0,1]$ follow the same uniform distribution. To establish the lower bound, we choose the distance between $\alpha_1$ and $\alpha_2$ as $\alpha_1-\alpha_2 = \frac{1}{4\sqrt{k}}$.
			
			Lower bounding $\left(\gamma_2-\gamma_1\right)^2$ will be broken down into two steps. First, we will show that $\gamma_2$ is greater than $\gamma_1$. Following this, we will use Taylor expansion to derive the lower bound of $\gamma_2$. The result is given in Lemma \ref{lemma:lower-bound-gamma}, and the proof is provided in Appendix M of the supplementary material \cite{supple}.
	
			\begin{lemma}
				With $\mathbb{P}_{\text{FD}}=\mathbb{P}_{\text{BD}}=\text{Uni}[0,1]$ and $\alpha_1-\alpha_2 =\frac{1}{4\sqrt{k}}$, we have the lower bound for $\gamma_2-\gamma_1$:
				\begin{equation}
					\gamma_2-\gamma_1\geq\frac{N_1}{7}\frac{1}{\sqrt{k}},
					\label{eq:lower-bound-gamma-diff}
				\end{equation}
				where $N_1$ is a constant.
				\label{lemma:lower-bound-gamma}
			\end{lemma}
					
			After lower bounding $(\gamma_2-\gamma_1)^2$, we continue to give a lower bound of $\mathbb{P}_1 \wedge \mathbb{P}_2$. Notice that :
			\begin{equation}
				\mathbb{P}_1^{\otimes k}\wedge\mathbb{P}_2^{\otimes k} = 1 - \frac{1}{2}\left|\mathbb{P}_1^{\otimes k}-\mathbb{P}_2^{\otimes k}\right|_1.
				\label{eq:relation-wedge}
			\end{equation}
			
			Then, it's sufficient to lower bound $\mathbb{P}_1^{\otimes k}\wedge\mathbb{P}_2^{\otimes k}$ as follows:
			\begin{align}
				\!\frac{1}{2}\!\vert\mathbb{P}_1^{\otimes k}\!-\!\mathbb{P}_2^{\otimes k}\vert_1
				\!\leq\!  \frac{1}{2}\!\sqrt{\!D_{\mathsf{KL}}(\mathbb{P}_1^{\otimes k}\Vert \mathbb{P}_2^{\otimes k})\!+\!D_{\mathsf{KL}}(\mathbb{P}_2^{\otimes k}\Vert \mathbb{P}_1^{\otimes k})},
			\end{align}
			where the inequality is by Pinsker's inequality: $\frac{1}{2}|\mathbb{P}_1^{\otimes k}\!-\!\mathbb{P}_2^{\otimes k}|_1\!\leq\!\sqrt{\frac{1}{2}D_{\mathsf{KL}}(\mathbb{P}_2^{\otimes k}||\mathbb{P}_1^{\otimes k})}$ and $\frac{1}{2}|\mathbb{P}_1^{\otimes k}\!-\!\mathbb{P}_2^{\otimes k}|_1\!\leq\!\sqrt{\frac{1}{2}D_{\mathsf{KL}}(\mathbb{P}_1^{\otimes k}||\mathbb{P}_2^{\otimes k})}$. Accroding to inequality $a+b \leq \sqrt{2(a^2+b^2)}$, we combine the KL divergence of the two distributions: $$\frac{1}{2}|\mathbb{P}_1^{\otimes k}-\mathbb{P}_2^{\otimes k}|_1\!\leq\!\frac{1}{2}\sqrt{2\!\times\!\frac{1}{2}\left(D_{\mathsf{KL}}(\mathbb{P}_1^{\otimes k}||\mathbb{P}_2^{\otimes k})\!+\!D_{\mathsf{KL}}(\mathbb{P}_2^{\otimes k}||\mathbb{P}_1^{\otimes k})\right)}.$$
			We further bound the terms in the square root as follows:
			\begin{align}
				&D_{\mathsf{KL}}(\mathbb{P}_1^{\otimes k}\Vert \mathbb{P}_2^{\otimes k})+D_{\mathsf{KL}}(\mathbb{P}_2^{\otimes k}\Vert \mathbb{P}_1^{\otimes k})\nonumber\\
				\overset{(a)}{=}&k\left(\log\left(\frac{1-\alpha_{1}}{1-\alpha_{2}}\right)-\left(1-\frac{1}{1-\alpha_{1}}\right)\log\frac{\alpha_{1}}{\alpha_{2}}\right)\nonumber\\
				&+k\left(\log\left(\frac{1-\alpha_{2}}{1-\alpha_{1}}\right)-\left(1-\frac{1}{1-\alpha_{2}}\right)\log\frac{\alpha_{2}}{\alpha_{1}}\right)\nonumber\\
				=&k\left(\frac{\alpha_1-\alpha_2}{(1-\alpha_1)(1-\alpha_2)}\right)\log\frac{\alpha_1}{\alpha_2}\nonumber\\
				\leq&k\left(\frac{\alpha_1-\alpha_2}{(1-\alpha_1)(1-\alpha_2)}\right)\left(\frac{\alpha_1}{\alpha_2}-1\right)\nonumber\\
				=&k\frac{(\alpha_1-\alpha_2)^2}{\alpha_2(1-\alpha_1)(1-\alpha_2)}\nonumber\\
				\overset{(b)}{=} & \frac{1}{16\alpha_2(1-\alpha_1)(1-\alpha_2)} \leq N_2.
			\end{align}
	
			By setting $\alpha_1 = \frac{1}{2}$, we have $\frac{1}{4}\leq\alpha_2 < \frac{1}{2}$. Therefore, we have $N_2 \leq 2$
			
			
			Equality (a) is derived from the computation of the KL divergence between geometric distributions. Equality (b) holds because $\alpha_1-\alpha_2=\frac{1}{\sqrt{k}}$. $N_2$ is a constant associated with $\alpha_1$ and $\alpha_2$. By selecting $\alpha_1$ and $\alpha_2$ carefully, we can obtain an upper bound for $\left|\mathbb{P}_1^{\otimes k}-\mathbb{P}_2^{\otimes k}\right|_1$:
			\begin{equation}
				\frac{1}{2}\left|\mathbb{P}_1^{\otimes k}-\mathbb{P}_2^{\otimes k}\right|_1\leq \frac{\sqrt{N_2}}{2}.
				\label{eq:bound-prob-L1}
			\end{equation}
			
			Then plugging \eqref{eq:bound-prob-L1} into \eqref{eq:relation-wedge}, we obtain the lower bound for $\mathbb{P}_1^{\otimes k}\wedge\mathbb{P}_2^{\otimes k}$:
			\begin{equation}
				\mathbb{P}_1^{\otimes k}\wedge\mathbb{P}_2^{\otimes k} \geq 1 - \frac{\sqrt{N_2}}{2}> 0.
				\label{eq:bound-proba}
			\end{equation}

			Finally, combining the lower bound of $\gamma_2-\gamma_1$, i.e., inequality \eqref{eq:lower-bound-gamma-diff} and $\mathbb{P}_1^{\otimes k}\wedge\mathbb{P}_2^{\otimes k}$, i.e., inequality \eqref{eq:bound-proba} into the Le Cam's inequality \eqref{eq:lecam-ineq}, we conclude that at the end of epoch $K$, the error for the estimator satisfies the lower bound:
			\begin{equation}
				\inf_{\hat{\gamma}}\sup_{\mathbb{P}}\mathbb{E}\left[\left(\hat{\gamma}\left(\mathcal{H}_K\right)-\gamma_{\mathbb{P}}^{\star}\right)^2\right]\geq\frac{N_1^2}{49}\left(1-\frac{\sqrt{N_2}}{2}\right)^2\!\cdot\! \frac{1}{K}.
			\end{equation}
	
			  \end{IEEEproof}
	
			


		\subsubsection{Proof of Equation \eqref{eq:minimax-2}}

		 \begin{IEEEproof}
	
		 We will first reformulate the cumulative AoI regret into an epoch-based AoI regret. The expected cumulative AoI up to epoch $K$ can be expressed as: 
			\begin{equation}
	\mathbb{E}\left[\int_0^{S_{K+1}}A(t)\text{d}t\right]=\mathbb{E}\left[\sum_{k=1}^{K}\left(Q_k+D^F_kL_{k-1}\right)\right].
			\end{equation}
			Therefore, we focus on deriving the bound for $Q_k+D^F_kL_{k-1}$ in each epoch $k$. Let $p_w=\text{Pr}(D^{\rm a}\leq \gamma^{\star}_{\mathbb{P}})$ be the probability of waiting for the first sample in each epoch.
			\begin{lemma}
				For any stationary waiting time selection function $w$, the expected reward $q$, epoch length $l$, and the probability of waiting $p_w$ satisfy the following inequality:
				\begin{equation}
					q\geq \mathbb{E}\left[D^{\rm v}\right]l + \gamma^{\star}_{\mathbb{P}}L+\frac{1}{2}p_w\left(l-\overline{L}^{\star}\right)^2.
				\end{equation}
				\label{lemm:bound-for-reward}
			\end{lemma}
			The proof for Lemma~\ref{lemm:bound-for-reward} is provided in Appendix N of the supplementary material \cite{supple}.
	
			With Lemma \ref{lemm:bound-for-reward}, in each epoch $k$, given historical information $\mathcal{H}^{\otimes k-1}$ and taking the expectation with respect to $\mathbb{P}$, the expected reward $Q_k$ and epoch length $L_k$ under any casual waiting time selection function $w$ satisfy:
			\begin{align}
				\mathbb{E}\left[Q_k\vert\mathcal{H}^{\otimes k-1}\right]\geq& \gamma^{\star}\mathbb{E}\left[L_k\vert\mathcal{H}^{\otimes k-1}\right]+\mathbb{E}\left[D^{\rm v}\right]\mathbb{E}\left[L_k\vert\mathcal{H}^{\otimes k-1}\right]\nonumber\\
				&+\frac{1}{2}p_w\left(\mathbb{E}\left[L_k\vert\mathcal{H}_{k-1}\right]-\overline{L}^{\star}\right)^2.
				\label{eq:relation-expect-condi}
			\end{align}
			Adding $\mathbb{E}\left[D^F_k L_{k-1}\vert \mathcal{H}^{\otimes k-1}\right]$ on both sides of \eqref{eq:relation-expect-condi}, the left-hand-side is the AoI accumulation in epoch $k$, i.e., $F_k=Q_k+D_k^F L_{k-1}$. Since the forward delay $D^F_k$ is independent of the historical data $\mathcal{H}^{\otimes k-1}$ and the previous epoch length $L_{k-1}$, we can express $\mathbb{E}\left[D^F_k L_{k-1}\vert \mathcal{H}^{\otimes k-1}\right]$ as $L_{k-1}\overline{D^F}$. We obtain a lower bound for the cumulative AoI in epoch $k$:
			\begin{align}
				&\mathbb{E}\left[Q_k+ D_k^F L_{k-1} \vert \mathcal{H}^{\otimes k-1}\right]  
				\geq \gamma^{\star}\mathbb{E}\left[L_k\vert\mathcal{H}^{\otimes k-1}\right]+\overline{D^F}L_{k-1}\nonumber\\
	   &\hspace{1cm}+\overline{D^{\rm v}}L_{k-1}
				+\frac{1}{2}p_w\left(\mathbb{E}\left[L_k\vert\mathcal{H}^{\otimes k-1}\right]-\overline{L}^{\star}\right)^2.
				\label{eq:relation-DF}
			\end{align}
			Next, we continue to derive the lower bound for the cumulative AoI from epoch 1 to $K$. Denote $l_k(\mathcal{H}_k)=\mathbb{E}\left
	  [L_k\vert\mathcal{H}^{\otimes k-1}=h_{k-1}\right]$ to be the expected epoch length obtained by function $w$, conditional on the historical transmission delays up to epoch $k-1$. Summing up \eqref{eq:relation-DF} from epoch $1$ to $K$ and take the expectation with respect to $\mathcal{H}^{\otimes k}$, we have:
			\begin{align}
				&\mathbb{E}\left[\sum_{k=1}^K\left(Q_k+D^F_kL_{k-1}\right)\right]\nonumber\\
				\geq& \left(\gamma^{\star}_{\mathbb{P}}+\overline{D^F}+\overline{D^{\rm v}}\right)\mathbb{E}\left[\sum_{k=1}^{K}L_{k}\right]\!-\!\left(M_{\text{ub}}\left(D^F_{\text{ub}}\!+\!D^B_{\text{ub}}\right)\!+\!W_{\text{ub}}\right)\nonumber\\
				&\cdot\left(\overline{D^F}+\overline{D^{\rm v}}\right) +\frac{1}{2}p_{w}\mathbb{E}\left[\sum_{k=1}^{K}\left(l_{k}(\mathcal{H}_{k-1})-\overline{L}^{\star}\right)^{2}\right] ,
				\label{eq:relation-summup}
			\end{align}
			where we define $L_0=0$ and use inequality $L_K\leq L_{\text{ub}}=M_{\text{ub}}(D^F_{\text{ub}}+D^B_{\text{ub}})+W_{\text{ub}}$.
			
			Recall that the optimal time-average AoI under delay distribution $\mathbb{P}$ is expressed by $\overline{A}_{w_{\mathbb{P}}^{\star}}=\gamma^{\star}_{\mathbb{P}}+\overline{D^F}+\overline{D^{\rm v}}$. Therefore, for any casual policy $w$, the cumulative AoI regret can be expressed as:
			\begin{align}
				&\inf_w \sup_{\mathbb{P}} \, \mathbb{E}\left[\int_0^{S_{K+1}}A(t)\text{d}t-S_{k+1}\overline{A}_{w_{\mathbb{P}}^{\star}}\right]\nonumber\\
				=&\inf_w \sup_{\mathbb{P}} \,\sum_{k=1}^{K}\mathbb{E}\left[Q_k+D^F_kL_{k-1}-L_{k}\overline{A}_{w_{\mathbb{P}}^{\star}}\right]\nonumber\\
			\geq & -      \left(\overline{D^F}+\overline{D^{\rm v}}\right)\left(M_{\text{ub}}(D^F_{\text{ub}}+D^B_{\text{ub}})+W_{\text{ub}}\right)\nonumber\\
				&+\inf_w \sup_{\mathbb{P}} \,\frac{1}{2}p_w \times \mathbb{E}\left[\sum_{k=1}^{K}(l_k(\mathcal{H}_{k-1})-\overline{L^{\star}})^2\right]\nonumber\\
			\geq & -\left(\overline{D^F}+\overline{D^{\rm v}}\right)\left(M_{\text{ub}}\left(D^F_{\text{ub}}+D^B_{\text{ub}}\right)+W_{\text{ub}}\right)\nonumber\\
				&+\sum_{k=1}^{K}\inf_w \max_{\mathbb{P}\in \{\mathbb{P}_1, \mathbb{P}_2\}} \,\frac{1}{2}p_w\left(\mathbb{P}\right) \times \mathbb{E}\left[\left(l_k(\mathcal{H}_{k-1})-\overline{L^{\star}}\right)^2\right]\nonumber\\
			\geq & -\left(\overline{D^F}+\overline{D^{\rm v}}\right)\left(M_{\text{ub}}\left(D^F_{\text{ub}}+D^B_{\text{ub}}\right)+W_{\text{ub}}\right)\nonumber\\
				&+\sum_{k=1}^{K}\frac12\underbrace{\min\{p_w(\mathbb{P}_1),p_w(\mathbb{P}_2)\}}_{=:H_1}\nonumber\\&\times\underbrace{\inf_{w}\max_{\mathbb{P}\in\{\mathbb{P}_1,\mathbb{P}_2\}}\mathbb{E}_{\mathcal{H}_{k-1}}\left[\left(\mathbb{E}[L_k|\mathcal{H}_{k-1}]-L^\star_\mathbb{P}\right)^2\right]}_{=:H_2}.
				\label{eq:converse-bound-K}
			\end{align}
			
			To establish the minimax lower bound of the cumulative AoI regret, we need to obtain the lower bound of terms $H_1$ and $H_2$, respectively. 
	  
			\textbf{For $H_1$}: Notice that we assume the same delay distributions $\mathbb{P}_{\text{FD}}=\mathbb{P}_{\text{BD}}=\text{Uni}[0,1]$ for $\mathbb{P}_1$ and $\mathbb{P}_2$. We can calculate the waiting probability as follows:
			\begin{align}
				p_w = & \min\left\{p_w(\mathbb{P}_1),p_w(\mathbb{P}_2)\right\}\nonumber\\
				= & \min \left\{\text{Pr}(D^{\rm a}\leq \gamma^{\star}_1), \text{Pr}(D^{\rm a}\leq \gamma^{\star}_2)\right\}\nonumber \\
				= &\text{Pr}(D^{\rm a}\leq \gamma^{\star}_1) 
				\geq \frac{8}{9}.\label{eq:bound-H1}
			\end{align}
	
			\textbf{For $H_2$}: The result is provided in Lemma \ref{lemm:bound-for-z}.
			\begin{lemma}
				For any mapping rule $l_{k+1} :\mathcal{H}^{\otimes {k}}\rightarrow \mathbb{R}$, we have:
				\begin{align}
					\inf_{l_{k+1}}\sup_{\mathbb{P}}\,&\mathbb{E}\left[\left(l_{k+1}(\mathcal{H}_{k})-L^\star_\mathbb{P}\right)^{2}\right]\geq\Omega\left(\frac{1}{k}\right).
					\label{eq:bound-for-z}
				\end{align}
				\label{lemm:bound-for-z}
			\end{lemma}
			The proof for Lemma 5 is provided in Appendix O of the supplementary material \cite{supple}. Plugging in the bound of waiting probability \eqref{eq:bound-H1} and epoch length \eqref{eq:bound-for-z}, we obtain the minimax bound for cumulative AoI regret:
			\begin{align}
	&\inf_{w}\sup_{\mathbb{P}}\,\mathbb{E}\left[\int_0^{S_{K+1}}A(t)\text{d}t\right]-\mathbb{E}[S_{K+1}]\overline{A}_{w^\star_\mathbb{P}} \nonumber\\
				\geq&-\left(M_{\text{ub}}(D^F_{\text{ub}}+D^B_{\text{ub}})+W_{\text{ub}}\right)\nonumber\\
	   &+\frac12 p_{w}\sum_{k=1}^{K}\inf_{l_{k}}\sup_{\mathbb{P}}(\mathbb{P})\times\mathbb{E}\left[\left(l_{k}(\mathcal{H}_{k-1})-L^\star_\mathbb{P}\right)^{2}\right] \nonumber\\
				\overset{(a)}{\geq}&\frac{4}{9}\cdot\Omega\left(\sum_{k=2}^{K}\frac{1}{k-1}\right)
				\overset{(b)}{\geq}\Omega\left(\ln K\right),
			\end{align}
			where inequality (a) is from inequality \eqref{eq:bound-H1} and \eqref{eq:bound-for-z} and inequality (b) is from $\sum_{k=1}^{K}\frac{1}{k} \geq \ln K$.
	
		\end{IEEEproof}

	\section{Conclusion \label{sec:conclusion}}
	In this paper, we studied a status update system where a sensor transmits status updates to a receiver through an unreliable channel with delayed feedback. We aimed to minimize the average AoI at the receiver while satisfying the sensor's sampling frequency constraint with unknown channel statistics. The problem was first reformulated into a stochastic approximation problem, and we proposed a Robbins-Monro-based algorithm that is capable of adaptively learning the AoI minimum sampling policy. Additionally, we enhanced the algorithm by incorporating momentum-based adjustments to reduce variance. Theoretical analysis demonstrates that the both the threshold $\gamma$ and cumulative age converge to the values under the optimal policy almost surely. Besides, the optimality gap of cumulative age decays with rate $\mathcal{O}\left(\ln K\right)$, and by Le Cam's two-point method, this gap matches the minimax order optimality. Simulation results validate the convergence and performance of the proposed algorithm.

	\bibliographystyle{IEEEtran}
	\bibliography{bibtex/bib/Stochastic_Approximation}

	\newpage
	\section*{Supplementary Material}
	\appendices
	\pagenumbering{arabic} 
\setcounter{page}{1}

	\section{Notations}
	We summarize main notations in the proof in Table~\ref{tab:notations}. 
	\begin{table}[h]
		\centering
		\caption{Notations}
		\begin{tabular}{p{1cm}<{\centering}|p{7cm}}
			\hline \hline
			Notations & Meaning \\
			\hline
			$D^F_{k,j}$ & The forward delay of the $j$-th sample of the $k$-th epoch. \\
			$D^B_{k,j}$ & The backward delay of the $j$-th sample of the $k$-th epoch. \\
			$M_k$ & The transmission times of epoch $k$.\\
			$w$ & The waiting time selection function.\\
			$D^{\rm a}_k$ & The actual delay in epoch $k$: $D^{\rm a}_k:=D^F_{k,1}+D^B_{k,1}$. \\
			$D^{\rm v}_k$ & The virtual delay after the first sample in epoch $k$: $D^{\rm v}_k:=\sum_{j=2}^{M_k}\left(D^F_{k,j}+D^B_{k,j}\right)$. \\
			$Q_k$ & $Q_k:=\frac{1}{2}(D^F_{k,1}+D^B_{k,1}+W_{k,1})^2$. \\
			$L_k$ & The duration of epoch $k$: $L_k:=D^F_{k,1}+D^B_{k,1}+W_{k,1}+D^{\rm v}_k$.\\
			$F_k$ & The AoI accumulation in $k$-th epoch, i.e., $F_k = \int^{S_{k+1,1}}_{S_{k,1}}A(t)\text{d}t$.\\
			$D$ & D is the total delay in an epoch, i.e., $D=D^F+D^B+D^{\rm v}$.\\
			$\overline{H}$ & The second moments of the delays. $\overline{H} = \overline{(D^F+D^B+D^{\rm v})^2}$, $\overline{H^F} =\overline{(D^F)^2} $, $\overline{H^B} =\overline{(D^B)^2} $, $\overline{H^{\prime}} =\overline{(D^{\rm v})^2} $.\\
			$\overline{A}_{w^{\star}_{\mathbb{P}}}$ & The expected time-averaged AoI using the optimal policy under distribution set $\mathbb{P}$. Notice that $\overline{A}_{w^{\star}_{\mathbb{P}}}$ = $\text{AoI}_{\text{opt}}$.\\
			$\mathbb{E}_k[\cdot]$ & The expectation of variable given historical observation $\mathcal{H}_{k-1}$. \\
			$\mathcal{L}$ & The Lagrange function associated with the optimization problem.\\
			$\delta \mathcal{L}$ & The Gateaux derivative of the Lagrange function.\\
			$D_{\mathsf{KL}}(\cdot||\cdot)$ & The KL divergence between two distributions.\\
			$\delta M_k$ & The martingale sequence depending on the context.\\   
			
			\hline
		\end{tabular}
		\label{tab:notations}
	\end{table}
	
	\section{Proof for Fractional Programming Reformulation\label{sec:reformulation}}
	
	\begin{IEEEproof}
		We first turn the problem from time-average computation into per-epoch computation and then transform it into a fractional programming.
		
		Notice that the AoI accumulation in the $k$-th epoch $F_k \triangleq \int_{S_{k,1}}^{S_{k+1,1}}A(t)\text{d}t$ can be computed by the area of a parallelogram and a triangle. Therefore, we can rewrite $F_k$ as follows:
		\begin{align}
			F_k =& D^F_{k,1} \times \sum_{j=1}^{M_{k-1}}(D^B_{k-1,j}+D^F_{k-1,j}+W_{k-1,j})\nonumber\\
			&+\frac{1}{2}\left[ \sum_{j=1}^{M_{k}}\left(D^B_{k,j}+D^F_{k,j}+W_{k,j}\right)\right]^2. 
			\label{eq:definition-F}
		\end{align}
		
		Next, we focus on stationary and deterministic policies that satisfy (6). The expected cumulative AoI in the $k$-th epoch can be computed by
		\begin{align}
			\mathbb{E}[F_k]
			=& \mathbb{E}\left[D^F_{k,1}\times \sum_{j=1}^{M_{k-1}}\left(D^B_{k-1,j}+D^F_{k-1,j}+W_{k-1,j}\right)\right.\nonumber\\
			&+\frac{1}{2}\left.\left[ \sum_{j=1}^{M_{k}}\left(D^B_{k,j}+D^F_{k,j}+W_{k,j}\right)\right]^2\right]\nonumber \\
			\overset{(a)}{=}&\mathbb{E}\left[D_{k, 1}^F\times\left(D_{k-1, 1}^B+D_{k-1, 1}^F+W_{k-1, 1}+D_{k-1}^{\rm v}\right)\right]\nonumber\\
			&+\frac{1}{2}\mathbb{E}\left[\left(D_{k, 1}^F+D_{k, 1}^B+W_{k, 1}+D_{k}^{\rm v}\right)^2\right]\nonumber \\
			\overset{(b)}{=}&\mathbb{E}\left[D^F\right]\left(\mathbb{E}\left[D^{\rm a}+w\right]+\mathbb{E}[D^{\rm v}]\right)+\frac{1}{2}\mathbb{E}\left[\left(D^{\rm a}+w\right)^2\right]\nonumber\\
			&+\frac{1}{2}\mathbb{E}\left[(D^{\rm v})^2\right]+\mathbb{E}\left[D^{\rm v}\right]\mathbb{E}\left[D^{\rm a}+w\right],
			\label{eq:cumulative-aoi}
		\end{align}
		where equality (a) is obtained because the additional virtual delay $D_k^{\rm v}=\sum_{j=2}^{M_k}(D_{k, j}^F+D_{k,j}^B)$ and $W_{k,j}=0, j \geq 2$; equality (b) is because $D_{k, 1}^F$ is independent of the delay distribution in epoch $k-1$, and that $D^{\rm v}_k$ is independent of $D_{k, 1}^B, D_{k, 1}^F$ and the waiting time $W_{k, 1}$. 
		
		Also, notice that the time interval in the $k$-th epoch can be expressed as $S_{k+1,1}-S_{k,1}=\sum_{j=1}^{M_{k}}(D^B_{k,j}+D^F_{k,j}+W_{k,j})$. $L_k$ can be rewritten as follows:
		\begin{align}
			L_k=\sum_{j=1}^{M_{k}}(D^B_{k,j}+D^F_{k,j}+W_{k,j}).
		\end{align}
		
		We can further simplify the expected length of $k$-th epoch as follows: 
		\begin{align}
			\mathbb{E}[L_k] & = \mathbb{E}\left[\sum_{j=1}^{M_{k}}(D_{k,j}^B+D_{k,j}^F+W_{k,j})\right] \nonumber\\
			& = \mathbb{E}\left[\sum_{j=1}^{M}(D^B_j+D^F_j+W_{j})\right]\nonumber\\
			& = \mathbb{E}[D^{\rm a}+w]+\mathbb{E}[D^{\rm v}].\label{eq:cumulative-length}
		\end{align}
		
		Finally, the average AoI in Problem~1 can be computed by:
		\begin{align}
			&\limsup_{T\rightarrow\infty}\frac{1}{T}\mathbb{E}\left[\int_0^TA(t)\text{d}t\right]\nonumber\\
			=&\limsup_{K\rightarrow\infty}\frac{\sum_{k=1}^K\mathbb{E}[F_k]}{\sum_{k=1}^K\mathbb{E}[L_k]}\nonumber\\
			=&\frac{\mathbb{E}\left[D^F\right]\left(\mathbb{E}\left[D^{\rm a}+w\right]+\mathbb{E}[D^{\rm v}]\right)}{\mathbb{E}[D^{\rm a}+w]+\mathbb{E}[D^{\rm v}]}\nonumber\\
			&+\frac{\frac{1}{2}\mathbb{E}\left[(D^{\rm a}+w)^2\right]+\frac{1}{2}\mathbb{E}[(D^{\rm v})^2]+\mathbb{E}[D^{\rm v}]\mathbb{E}[D^{\rm a}+w]}{\mathbb{E}[D^{\rm a}+w]+\mathbb{E}[D^{\rm v}]}\nonumber\\
			=&\mathbb{E}[D^F]+\mathbb{E}[D^{\rm v}]\nonumber\\
			&+\frac{\frac{1}{2}\mathbb{E}\left[(D^{\rm a}+w)^2\right]+\frac{1}{2}\mathbb{E}[(D^{\rm v})^2]-\mathbb{E}[D^{\rm v}]^2}{\mathbb{E}[D^{\rm a}+w]+\mathbb{E}[D^{\rm v}]}.
		\end{align}
		
	\end{IEEEproof}
	
	\section{Proof for proposition \ref{prop:opt-pi}\label{sec:prop-opt-pi}}
	\begin{IEEEproof}
		
		The Lagrange function is as follows:
		\begin{align}
			\mathcal{L}(\nu, w, \mu):=& \frac{1}{2}\mathbb{E}\left[\left(D^{\rm a}+w(D^{\rm a})\right)^2\right]+\frac{1}{2}\mathbb{E}[{D^{\rm v}}^2]\nonumber\\
			&-(\text{AoI}_{\text{opt}}-\mathbb{E}[D^F]-\mathbb{E}[D^{\rm v}])\mathbb{E}[D^{\rm a}+w(D^{\rm a})]\nonumber\\
			&-(\text{AoI}_{\text{opt}}-\mathbb{E}[D^{\rm v}])\mathbb{E}[D^{\rm v}]+\mathbb{E}\left[\mu(D^{\rm a}) w(D^{\rm a})\right]\nonumber\\
			&+\nu\left(\frac{\mathbb{E}[M]}{f_{\text{max} }}-\mathbb{E}[D^{\rm a}+w(D^{\rm a})]-\mathbb{E}[D^{\rm v}]\right).
			\label{eq:lagrange-NB-beta-app}
		\end{align}

		
		KKT condition remains valid for Lebesgue space $L^2(\cdot)$. Therefore, a vector $(w, \nu, \theta)$ is an optimal solution if it satisfies the KKT conditions given as follows
		\begin{subequations}
			\begin{align}
				w^{\star} = &\arg \min_w \mathcal{L}(\nu, w, \mu),\\
				&\nu \geq 0,\\
				&w(D^{\rm a})\geq 0,\\
				&\mathbb{E}[D^{\rm a}+w(D^{\rm a})]-\mathbb{E}[D^{\rm v}] \geq \frac{\mathbb{E}[M]}{f_{\text{max} }},\\
				&\nu\left(\frac{\mathbb{E}[M]}{f_{\text{max} }}-\mathbb{E}[D^{\rm a}+w(D^{\rm a})]-\mathbb{E}[D^{\rm v}]\right) = 0,\label{eq:CS-1-prop}\\
				&w(D^{\rm a}) \mu(D^{\rm a})  = 0, \forall D^{\rm a} \geq 0,\label{eq:CS-2-prop}
			\end{align}
		\end{subequations}
		where the equalities \eqref{eq:CS-1-prop} and \eqref{eq:CS-2-prop} are from Complete Slackness (CS) conditions.
		
		Then, we solve the KKT conditions with the calculus of variations. For fixed vector $(\nu, \mu)$, the Gateaux derivative of the Lagrange function \eqref{eq:lagrange-NB-beta-app} in the direction of $w\in L^2$ is denoted by $\delta\mathcal{L}(w;\nu, \mu, \theta)$:
		\begin{align}
			\delta\mathcal{L}(w;\nu, \mu, \theta)
			\triangleq&\lim_{\epsilon \rightarrow 0}\frac{\mathcal{L}(w+\epsilon\theta,\nu, \mu) - \mathcal{L}(w, \nu, \mu)}{\epsilon}\nonumber\\
			=&\mathbb{E}\left[\left(D^{\rm a}+w(D^{\rm a})\right.\right.\nonumber\\
			&\left.\left.-(\text{AoI}_{\text{opt}}-\mathbb{E}[D^F]-\mathbb{E}[D^{\rm v}])-\nu+\mu\right)\theta\right].
			\label{eq:gate-lagrange}
		\end{align}
		
		Then, $w(\cdot)$ is an optimal solution if and only if
		\begin{align}
			\delta\mathcal{L}(w;\nu, \mu, \theta) \geq 0, \forall \theta \in L^2.
		\end{align}
		
		Since $\delta\mathcal{L}(w;\nu, \mu, \theta) = -\delta\mathcal{L}(w;\nu, \mu, -\theta) = 0$, we have the condition for the optimal solution:
		\begin{equation}
			\delta\mathcal{L}(w;\nu, \mu, \theta) = 0, \forall \theta \in L^2,
			\label{eq:KKT-condition-prop}
		\end{equation}
		
		
		Notice that $\theta$ is arbitrary. Plugging Gateaux derivative \eqref{eq:gate-lagrange} into the KKT conditions \eqref{eq:KKT-condition-prop}, we obtain:
		\begin{equation}
			D^{\rm a}+w(D^{\rm a})-(\text{AoI}_{\text{opt}}-\mathbb{E}[D^F]-\mathbb{E}[D^{\rm v}])-\nu+\mu = 0.
		\end{equation}
		Considering the CS conditions \eqref{eq:CS-1-prop} and \eqref{eq:CS-2-prop}, the optimal policy $w^{\star}$ can be obtained as follows:
		\begin{equation}
			w^{\star}(D^{\rm a}) = \left(\nu + \left(\text{AoI}_{\text{opt}}-\mathbb{E}[D^F]-\mathbb{E}[D^{\rm v}]\right) - D^{\rm a}\right)^{+}.
		\end{equation}
	\end{IEEEproof}

	\section{Proof for Lemma \ref{lemma:gamma-bound}\label{appd:proof-gamma-bound}}
	\begin{IEEEproof}
		
		First, we derive the lower bound for $\gamma^{\star}$ using the bounds of the  delays. 
		\begin{align}
			\gamma^{\star}
			=&\frac{\mathbb{E}\left[\left(D^B+D^F+w+D^{\rm v}\right)^2\right]}{2\mathbb{E}\left[D^B+D^F+w+D^{\rm v}\right]} - \mathbb{E}[D^{\rm v}]\nonumber\\
			\overset{(a)}{\geq}&\frac{1}{2}\frac{\mathbb{E}\left[\left(D^B+D^F+w+D^{\rm v}\right)\right]^2}{\mathbb{E}\left[D^B+D^F+w+D^{\rm v}\right]} - \mathbb{E}[D^{\rm v}]\nonumber\\
			=&\frac{1}{2}\mathbb{E}\left[D^B+D^F+w-D^{\rm v}\right]\nonumber\\
			\overset{(b)}{\geq}&\frac{1}{2}\mathbb{E}\left[D^B+D^F-D^{\rm v}\right]
			\overset{(c)}{\geq}\frac{1}{2}(\overline{D^F}_{\text{lb}}+\overline{D^B}_{\text{lb}}-\overline{D^{\rm v}}_{\text{ub}}),
		\end{align}
		where inequality (a) is from Jensen's inequality; inequality (b) is because $0\leq w\leq W_{\text{ub}}$ and inequality (c) is obtained by Assumption 1.
		
		Notice that $\gamma^{\star} \geq 0$. Then we obtain the lower bound for $\gamma^{\star}$:
		\begin{equation}
			\gamma_{\text{lb}} = \max\{\frac{1}{2}(\overline{D^F}_{\text{lb}}+\overline{D^B}_{\text{lb}}-\overline{D^{\rm v}}_{\text{ub}}), 0\}.
		\end{equation}
		
		Next, we will utilize the constant wait policy $w_{\text{const}}$ to obtain the upper bound of $\gamma^{\star}$. Consider a policy that chooses waiting time $w_{\text{const}} = \frac{1}{f_{\text{max}}}$. Then, the expected average AoI of the constant wait policy can be computed by:
		\begin{align}
			\overline{A}_{w_{\text{const}}}
			=&\mathbb{E}[D^F]+\frac{\mathbb{E}\left[\left(D^B+D^F+w_{\text{const}}+D^{\rm v}\right)^2\right]}{2\mathbb{E}\left[D^B+D^F+w_{\text{const}}+D^{\rm v}\right]}\nonumber\\
			= & \frac{\frac{1}{2}\mathbb{E}[(D^F\!+\!D^B\!+\!D^{\rm v})^2]\!+\!\mathbb{E}[D^F\!+\!D^B\!+\!D^{\rm v}]\frac{1}{f_{\text{max}}} \!+\! \frac{1}{f_{\text{max}}^2}}{\mathbb{E}[D^F+D^B+D^{\rm v}]+\frac{1}{f_{\text{max}}}}\nonumber\\
			&+\mathbb{E}[D^F].
		\end{align}
		Since the constant wait policy is not the optimal policy, the expected average AoI of the constant wait policy will be greater than the optimal AoI, expressed as $\text{AoI}_{\text{opt}}\leq\overline{A}_{w_{\text{const}}}$. Leveraging this property, we obtain the upper bound for $\gamma^{\star}$ as follows:
		\begin{align}
			\gamma^{\star} 
			=& \beta^{\star} - \mathbb{E}[D^{\rm v}]\nonumber\\
			=&\text{AoI}_{\text{opt}} - \mathbb{E}[D^F]- \mathbb{E}[D^{\rm v}]\nonumber\\
			\leq&\frac{\frac{1}{2}\mathbb{E}[(D^F\!+\!D^B\!+\!D^{\rm v})^2]\!+\!\mathbb{E}[D^F\!+\!D^B\!+\!D^{\rm v}]\frac{1}{f_{\text{max}}} \!+\! \frac{1}{f_{\text{max}}^2}}{\mathbb{E}[D^F+D^B+D^{\rm v}]+\frac{1}{f_{\text{max}}}}\nonumber\\
			&-\mathbb{E}[D^{\rm v}]\nonumber\\
			\leq&\frac{\frac{1}{2}\overline{H}_{\text{ub}}+\overline{D}_{\text{ub}}\frac{1}{f_{\text{max}}} + \frac{1}{f_{\text{max}}^2}}{\overline{D}_{\text{lb}}+\frac{1}{f_{\text{max}}}}-\overline{D^{\rm v}}_{\text{lb}}=\gamma_{\text{ub}},
			\label{eq:gamma-upp}
		\end{align}
		where we denote the expectation and the second moment of the delays as $\overline{D}_{\text{ub}} = \overline{D^F}_{\text{ub}}+\overline{D^B}_{\text{ub}}+\overline{D^{\rm v}}_{\text{ub}}$, $\overline{H}_{\text{ub}} = \overline{(D^F+D^B+D^{\rm v})^2}_{\text{ub}}$, respectively.
		
	\end{IEEEproof}

	\section{Proof of Claim \ref{sec:claim}}\label{sec:pf-claim}
	\begin{IEEEproof}
		
		We will prove each condition in Claim~1 respectively. 
		
		\textbf{(1.1)} We will prove claim (1.1) by directing upper bound $\mathbb{E}[Y_k^2]$ for each epoch $k$. Notice that when there is no sampling constraint, $\nu_k\equiv 0$ and $\mathbb{E}[Y_k^2]$ can be upper bounded as follows:
		\begin{align}
			&\mathbb{E}[Y_k^2]\nonumber\\
			\overset{(a)}{=}&\mathbb{E}\Big[\Big(\frac{1}{2}\max\{D_k^{\rm a}, \gamma_k\}^2\nonumber\\
			&-\gamma_k\left(\max\{D_k^{\rm a}, \gamma_k\}+D_k^{\rm v}\right)+\frac{1}{2}m_k-\mu_k^2\Big)^2\Big]\nonumber\\
			\overset{(b)}{\leq}&2\mathbb{E}\left[\left(\frac{1}{2}\max\{D_k^{\rm a}, \gamma_k\}^2-\gamma_k(\max\{D_k^{\rm a}, \gamma_k\}+D_k^{\rm v})\right)^2\right]\nonumber\\
			&+2\mathbb{E}\left[\left(\frac{1}{2}m_k-\mu_k^2\right)^2\right]\nonumber\\
			\leq&\frac{1}{2}\left(\mathbb{E}[(D_k^{\rm a})^4]+\gamma_{\text{ub}}^4\right)+2\gamma_{\text{ub}}^2(\gamma_{\text{ub}}^2+\mathbb{E}[(D^{\rm v})^2])\nonumber\\
			&+\frac{1}{2}\mathbb{E}[m_k^2]+2\mathbb{E}[\mu_k^4]\nonumber\\
			\overset{(c)}{\leq}&B+\frac{5}{2}\gamma_{\text{ub}}^4+2\gamma_{\text{ub}}^2\frac{\alpha^2}{1-\alpha}\sqrt{B}+\frac{1}{2}B.
		\end{align}
		where equation (a) is obtained from (18); inequality $(b)$ is obtained because $\mathbb{E}[(X+Y)^2]\leq2\mathbb{E}[X^2]+2\mathbb{E}[Y^2]$; inequality $(c)$ is obtained because $D_k^{\rm a}=D_{k, 1}^F+D_{k,1}^B$ by definition and thus $\mathbb{E}[(D_k^{\rm a})^4]\leq 2B$
		
		\textbf{(1.2)} We will prove sequence $\{\delta M_{k,1}\}$ and $\{\delta M_{k,2}\}$ are martingales so that $\mathbb{E}_k[\delta M_{k, 1}]=0$ and $\mathbb{E}_k[\delta M_{k, 2}]=0$.  Notice that the transmission delay $D_k^{\rm a}$ and $D_k^{\rm v}$ are independent in each frame $k$, and $\overline{g}_0(\gamma_k)=\mathbb{E}[g_0(\gamma_k; D^{\rm a}, D^{\rm v})]$ by definition from (15). Therefore, 
		\begin{equation}
			\mathbb{E}_k[\delta M_{k, 1}]=\mathbb{E}_k[g_0(\gamma_k; D^{\rm a}, D^{\rm v})]-\overline{g}_0(\gamma_k)=0,
		\end{equation}
		which shows $\delta M_{k, 1}$ is a martingale sequence. 
		
		We will then show $\delta M_{k,2}$ is a martingale sequence as well. By plugging the updates of $m_k$ and $\mu_k$ from (17b) and (17a) into the definition, $\delta M_{k, 2}$ can be compute as follows:
		\begin{equation}
			\delta M_{k, 2}=\frac{1}{k}(D_{k}^{\rm v}+\frac{1}{2}{D_k^{\rm v}}^2)-\frac{1}{k}N. 
		\end{equation}
		
		As $D_k^{\rm v}$ is independent in each slot, and $N=\frac{1}{2}\mathbb{E}[{D^{\rm v}}^2]-\mathbb{E}[D^{\rm v}]^2$ by definition, we have $\mathbb{E}[\delta M_{k, 2}]$ is a martingale sequence. 
		
		\textbf{(1.3)} Notice that function $g_0(\gamma; D^{\rm a}, D^{\rm v})$ is continuous according to the definition in (14). As $\overline{g}_0(\gamma)=\mathbb{E}[g_0(\gamma; D^{\rm a}, D^{\rm v})]$ by definition, function $\overline{g}_0(\gamma)$ is thus continuous. 
		
		\textbf{(1.4)} The selection of stepsizes in (19) suggests:
		\begin{equation}
			\sum_k\eta_k\leq\sum_{s=2}^\infty\frac{1}{s^2}\leq\int_{s=1}^\infty\frac{1}{s^2}\text{d}s<\infty. 
		\end{equation}
		
		\textbf{(1.5)} Notice that $m_k$ and $\mu_k$ is the estimation of $\mathbb{E}[{D^{\rm v}}^2]$ and $\mathbb{E}[D^{\rm v}]$ from i.i.d samples $\{D_1^{\rm v}, \cdots, D_k^{\rm v}\}$, therefore $m_k\overset{\text{a.s}}{\rightarrow}\mathbb{E}[(D^{\rm v})^2]$ and $\mu_k\overset{{\rm a.s.}}{\rightarrow}\mathbb{E}[D^{\rm v}]$. According to (13), $N=\frac{1}{2}\mathbb{E}[(D^{\rm})^2]-\mathbb{E}[D^{\rm v}]^2$, therefore, $b_k=\frac{k-1}{k}\left(\frac{1}{2}m_k-\mu_k^2-N\right)$ converges to 0 almost surely by law of large numbers. Recall that the stepsize $\eta_k=1/k$, therefore, sequence $\sum_k\eta_kb_k<\infty$ almost surely.  
		
	\end{IEEEproof}

	\section{Proof for Claim \ref{sec:claim-2}\label{appd:proof-claim-2}}
	\begin{IEEEproof}
		
		We will provide the proof for each condition in Claim 2 respectively.
		\begin{itemize}
			\item [\textbf{(2.1)}] In each epoch $k$, the delay $D^{\rm a}, D^{\rm v}$ and the threshold $\gamma_k$ are bounded. Therefore, $\theta_k$ is bounded and $\sup_k \mathbb{E}[| Y_k| ]$ is bounded.
			\item [\textbf{(2.2)}] Function $\overline{f}(\theta, \gamma)$ is continuous in $\theta$ by definition.
			\item [\textbf{(2.3)}] The difference between $\overline{f}(\theta,\gamma)$ and $\overline{f}(\theta,\gamma^{\star})$ can be bounded by
			\begin{align}
				& \mid \overline{f}(\theta,\gamma)-\overline{f}(\theta,\gamma^{\star})\mid\nonumber\\
				=&\mid \overline{g}(\gamma)-\overline{g}(\gamma^{\star})\mid \nonumber\\
				\leq&(\gamma-\gamma^{\star})^2 + \mid \gamma-\gamma^{\star} \mid \left(\overline{D^{\rm a}}+\overline{D^{\rm v}}\right)
			\end{align}
			
			According to \eqref{eq:theo-1}, we have $| \gamma_k - \gamma^{\star}|=\mathcal{O}(k^{-1/2})$. Then we have the limit for all $\theta$:
			\begin{align}
				&\lim_{k\to\infty}\Pr\left(\sup_{j\geq k}\left|\sum_{i=k}^{j}\epsilon_i\left(f(\theta,\gamma_i)-\overline{f}(\theta, \gamma^{\star})\right)\right|\geq\mu\right)\nonumber\\
				\leq & \frac{\mathbb{E}\left[\sup_{j\geq k}\left|\sum_{i=k}^j \epsilon_k\left(\overline{f}\left(\theta,\gamma_k\right)-\overline{f}\left(\theta,\gamma^{\star}\right)\right)\right|\right]}{\mu}\nonumber\\
				\leq &\frac{1}{\mu} \mathbb{E}\left[\sum_{i=k}^{\infty}\epsilon_i\cdot \left|\overline{f}(\theta,\gamma_k)-\overline{f}(\theta, \gamma^{\star})\right|\right]\nonumber\\
				\leq & \frac{1}{\mu}\mathbb{E}\left[\sum_{i=k}^{\infty}\frac{1}{i^{-3/2}}\right]=\mathcal{O}(k^{-1/2}).
				\label{eq:claim-2-3}
			\end{align}
			
			Taking the limit of both sides of inequality \eqref{eq:claim-2-3}, and recall $m(k)=\lfloor\exp(k)\rfloor$, we have 
			\begin{align}
				&\lim_{k\to\infty}\Pr\left(\sup_{j\geq m(k)}\left|\sum_{i=m(k)}^{j}\!\epsilon_i\!\left(f(\theta,\gamma_i)\!-\!\overline{f}(\theta, \gamma^{\star})\right)\right|\geq\!\mu\!\right)\nonumber\\
				\leq & \lim_{k\rightarrow \infty} \frac{2}{\mu} \frac{1}{\sqrt{\exp(k)-1}}=0.
			\end{align}
			
			\item [\textbf{(2.4)}] Given historical information $\mathcal{H}_{k-1}$, the martingale sequence $\delta M_k$ only depends on delay $D^{\rm a}_k, D^{\rm v}_k$ and has zero mean. Since $\gamma_k$ is upper bounded, delay $D^{\rm a},D^{\rm v}$ is second order bounded, $Y_k$ is bounded and the difference sequence $\delta M_k$ is second order bounded. Therefore, the sequence $M_k:=\sum_{k'=1}^k \epsilon_k \delta M_{k'}$ is also a martingale sequence. According to \cite[Chapter 5, Eq. (2.6)]{yu1997assouad}, for each $\mu >0$, we have 
			\begin{align}
				&\lim_{k\rightarrow \infty}\text{Pr}\left(\sup_{j\geq k}\left| \sum_{i=k}^j\epsilon_i\delta M_i\right|\geq \mu\right)\nonumber\\
				=&\lim_{k\rightarrow \infty}\text{Pr}\left(\sup_{j\geq k}\left| M_j-M_k\right|\geq \mu\right)=0.
			\end{align}
			
			\item [\textbf{(2.5)}] $\beta_{k,1}$ and $\beta_{k,2}$ can be viewed as two bias terms in the recursive form. Through union bound, we have 
			\begin{align}
				&\lim_{k\rightarrow \infty}\text{Pr}\left(\sup_{j\geq k}\left| \sum_{i=k}^j\epsilon_i(\beta_{k,1}+\beta_{k,2})\right|\geq \mu\right)\nonumber\\
				\leq&\lim_{k\rightarrow \infty}\text{Pr}\left(\sup_{j\geq k}\left| \sum_{i=k}^j\epsilon_i\beta_{k,1}\right|\geq \frac{\mu}{2}\right)\nonumber\\
				&+\lim_{k\rightarrow \infty}\text{Pr}\left(\sup_{j\geq k}\left| \sum_{i=k}^j\epsilon_i\beta_{k,2}\right|\geq \frac{\mu}{2}\right).
				\label{eq:union-bound}
			\end{align}
			
			The first term in \eqref{eq:union-bound} can be upper bounded as follows:
			\begin{align}
				&\lim_{k\rightarrow \infty}\text{Pr}\left(\sup_{j\geq k}\left| \sum_{i=k}^j\epsilon_i\beta_{k,1}\right|\geq \frac{\mu}{2}\right)\nonumber\\
				\leq & \frac{\mathbb{E}\left[\sup_{j\geq k}\left|\sum_{i=k}^j \epsilon_i \beta_{i,1}\right|\right]}{\mu/2}\!\leq  \!\frac{2}{\mu} \mathbb{E}\left[\sum_{i=k}^{\infty}\frac{1}{i}\left|\beta_{i,1}\right|\right].
			\end{align}
			The expectation of $\beta_{k,1}$ can be upper bounded as follows:
			\begin{align}
				\mathbb{E}[\beta_{k,1}]=&\mathbb{E}\left[\overline{D^F} (\mathbb{E}\left[L_{k-1}|\mathcal{H}_{k-1}\right]-l(\gamma_k))\right]\nonumber\\
				=&\mathbb{E}\left[l(\gamma_{k-1}) - l(\gamma_k)\right]\nonumber\\
				=&\mathbb{E}\left[l(\gamma_{k-1}) - l(\gamma^{\star})- \left(l(\gamma_k)- l(\gamma^{\star})\right)\right]\nonumber\\
				\leq&\mathbb{E}\left[ |\gamma_{k-1}-\gamma^{\star}|+|\gamma_{k}-\gamma^{\star}|\right]\nonumber\\
				=&\mathcal{O}(k^{-1/2}).
			\end{align}
			
			Therefore, we have 
			\begin{align}
				&\lim_{k\rightarrow \infty}\text{Pr}\left(\sup_{j\geq k}\left| \sum_{i=m(k)}^j\epsilon_i\beta_{k,1}\right|\geq \frac{\mu}{2}\right)\nonumber\\
				\leq & \lim_{k\rightarrow \infty}\frac{2}{\mu} \mathbb{E}\left[\sum_{i=k}^{\infty}i^{-3/2}\right]\nonumber\\
				\leq & \lim_{k\rightarrow \infty}\frac{2}{\mu} \mathcal{O}(\frac{1}{\sqrt{k}})
				=0.
			\end{align}
			
			Next, we move to bound the second part $\beta_{k,2}=(\gamma_k-\gamma^{\star})l(\gamma_k)$. $l(\gamma_k)$ is upper bounded since $\gamma_k$ and delays are upper bounded. $|\gamma_k-\gamma^{\star}|=\mathcal{O}(k^{-1/2})$. Therefore, through similar deduction as $\beta_{k,1}$, we have:
			\begin{align}
				&\text{Pr}\left(\sup_{j\geq k}\left| \sum_{i=m(k)}^j\epsilon_i\beta_{k,2}\right|\geq \frac{\mu}{2}\right)\nonumber\\
				\leq & \frac{2}{\mu} \mathbb{E}\left[\sum_{i=k}^{\infty}i^{-3/2}\right]
				=\mathcal{O}(k^{-1/2}).
			\end{align}
			
			Then we obtain the result:
			\begin{align}
				\lim_{k\rightarrow \infty}\text{Pr}\left(\sup_{j\geq k}\left| \sum_{i=m(k)}^j\epsilon_i\beta_{k,2}\right|\geq \frac{\mu}{2}\right)=0.
			\end{align}

			\item [\textbf{(2.6)}] Since the delays and waiting time are upper bounded, function $f$ is uniformly bounded for $\theta \in [0, 2L_{\text{ub}}^2]$, $\gamma\in [\gamma_{\text{lb}}, \gamma_{\text{ub}}]$.
			
			\item [\textbf{(2.7)}] According to the definition of $f$, for each $\gamma$ we have:
			\begin{align}
				|f(\theta_1, \gamma)-f(\theta_2, \gamma)|=|\theta_1-\theta_2|,
			\end{align}
			and $\lim_{|\theta_1-\theta_2| \rightarrow 0}|f(\theta_1, \gamma)-f(\theta_2, \gamma)|=0$.
			\item [\textbf{(2.8)}] Sequence $\frac{1}{k}$ satisfies $\sum_{k'=1}^{\infty}\frac{1}{k'}=\infty$.
		\end{itemize}
		
	\end{IEEEproof}

	

	\section{Proof for Lemma \ref{lemm:property-lyapunov-function}\label{appd:proof-lyapunov}}
	\begin{IEEEproof}
		
		Denote $\overline{L}^{\star}:=\mathbb{E}\left[D^{\rm a}+w^{\star}+D^{\rm v}\right]$ and $\overline{Q}^{\star}:=\frac{1}{2}\mathbb{E}\left[\max \{D^{\rm a}, \gamma^{\star} \}^2\right]$ to be the expected epoch length and the epoch reward when the optimal waiting time selection function $w^{\star}$ is used. To facilitate the proof, we first establish the connection between the epoch length, epoch reward, and the threshold $\gamma_k$ in Lemma \ref{lemma:bound-for-Q-L} and Lemma \ref{lemma:bound-for-gamma}.
		
		\begin{lemma}
			The expected epoch length $\mathbb{E}\left[L_k|\gamma_k\right]$ and $\mathbb{E}\left[Q_k| \gamma_k\right]$ in epoch $k$ satisfy:
			\begin{subequations}
				\begin{align}
					\mathbb{E}\left[Q_k - \gamma_k L_k + \frac{1}{2}m_k -\mu_k^2| \gamma_k\right] &\leq \left(\gamma^{\star}-\gamma_k\right)\overline{L}^{\star},
					\label{eq:reward-bound-1}
					\\
					\mathbb{E}\left[Q_k - \gamma_k L_k + \frac{1}{2}m_k -\mu_k^2| \gamma_k\right] &\leq - \left(\gamma^{\star}-\gamma_k\right)\mathbb{E}\left[L_k - \overline{L}^{\star}| \gamma_k\right].
					\label{eq:reward-bound-2}
				\end{align}
			\end{subequations}
			\label{lemma:bound-for-Q-L}
		\end{lemma}
		
		Proofs for Lemma \ref{lemma:bound-for-Q-L} is provided in Appendix \ref{sec:proof-bound-for-Q-L}. For simplicity, we denote $Q = \frac{1}{2}\max\{D^{\rm a}, \gamma\}$, $L=D^{\rm a}+D^{\rm v}+W$. Considering different values of $\gamma$, the analysis will be divided into two cases:
		\begin{itemize}
			\item If $\gamma -\gamma^{\star} > 0$, through \eqref{eq:reward-bound-1}, we have
			\begin{align}
				&\left(\gamma -\gamma^{\star}\right)\mathbb{E}\left[Q - \gamma L + \frac{1}{2}m -\mu^2\right] \nonumber\\
				\leq & \left(\gamma -\gamma^{\star}\right)\left(\gamma^{\star} - \gamma\right)\overline{L}^{\star} \nonumber \\
				\leq & -\left(\gamma -\gamma^{\star}\right)^2\left(\overline{D^{\rm a}} + \overline{D^{\rm v}}\right), 
				\label{eq:gamma-ineq-s-1}
			\end{align}
			where the last inequality is because $\overline{L}^{\star} \geq \overline{D^{\rm a}} + \overline{D^{\rm v}}$.
			
			\item If $\gamma - \gamma^{\star} \leq 0$, through \eqref{eq:reward-bound-2}, we have
			\begin{align}
				&\left(\gamma -\gamma^{\star}\right)\mathbb{E}\left[Q - \gamma L +\frac{1}{2}m -\mu^2\right] \nonumber\\
				= & \left(\gamma -\gamma^{\star}\right)\mathbb{E}\left[Q - \gamma^{\star} L + \frac{1}{2}m -\mu^2|\gamma\right] \nonumber \\
				&- \left(\gamma -\gamma^{\star}\right)^2\mathbb{E}\left[L\right]\nonumber \\
				\overset{(a)}{\leq} & \left(\gamma -\gamma^{\star}\right)\left(\overline{Q}^{\star} - \gamma^{\star} \overline{L}^{\star} - \mathbb{E}\left[\frac{1}{2}m -\mu^2\right]\right) \nonumber \\
				&- \left(\gamma_k -\gamma^{\star}\right)^2\mathbb{E}\left[L\right]\nonumber \\
				=&- \left(\gamma_k -\gamma^{\star}\right)^2\mathbb{E}\left[L\right]\nonumber\\
				\overset{(b)}{\leq}&- \left(\gamma_k -\gamma^{\star}\right)^2\left(\overline{D^{\rm a}} + \overline{D^{\rm v}}\right).
				\label{eq:gamma-ineq-s-2} 
			\end{align}
			Inequality (a) is obtained from the inequality $\mathbb{E}\left[Q_k - \gamma^{\star} L_k + \frac{1}{2}m_k -\mu_k^2|\gamma_k\right] \geq \overline{Q}^{\star} - \gamma^{\star} \overline{L}^{\star} + \frac{1}{2}\mathbb{E}\left[(D^{\rm v})^2\right]-\mathbb{E}\left[D^{\rm v}\right]^2\overline{L}^{\star} = 0$. Inequality (b) is because $\mathbb{E}\left[L_k|\gamma_k\right] \geq \overline{D^{\rm a}} + \overline{D^{\rm v}}$.
		\end{itemize}
	\end{IEEEproof}
	
	\section{Proof for Lemma 7\label{sec:proof-bound-for-Q-L}}
	\begin{IEEEproof}
		
		Notice that in each epoch $k$, the waiting time $w$ is selected to minimize the Lagrange function. Then we have:
		\begin{align}
			&\mathbb{E}\left[Q_k - \gamma_k L_k + \frac{1}{2}m_k -\mu_k^2|\gamma_k\right] \nonumber\\
			\overset{(a)}{\leq} & \overline{Q}^{\star} - \gamma_k \overline{L}^{\star} + \frac{1}{2}\mathbb{E}\left[(D^{\rm v})^2\right] -\mathbb{E}\left[D^{\rm v}\right]^2 \nonumber\\
			=&\overline{Q}^{\star} -\gamma^{\star} \overline{L}^{\star} + \gamma^{\star} \overline{L}^{\star} - \gamma_k \overline{L}^{\star} + \frac{1}{2}\mathbb{E}\left[(D^{\rm v})^2\right] -\mathbb{E}\left[D^{\rm v}\right]^2 \nonumber\\
			\overset{(b)}{=}&\gamma^{\star} \overline{L}^{\star} -\gamma_k \overline{L}^{\star},
			\label{eq:lemma-2-1}
		\end{align}
		where inequality (a) is because the waiting time $w_k$ used in epoch $k$ minimizes the Lagrange function and $D^{\rm v}_{k-1}$ is independent of $L_k$. Equality (b) is obtained because under the optimal policy we have $ \overline{Q}^{\star}-\gamma^{\star} \overline{L}^{\star} +\frac{1}{2}\mathbb{E}\left[(D^{\rm v})^2\right] - \mathbb{E}\left[D^{\rm v}\right]^2 = 0.$ Then the first inequality of Lemma \ref{lemma:bound-for-Q-L} has been proved. 
		
		For the second inequality, adding $\left(\gamma_k - \gamma^{\star}\right)\mathbb{E}\left[L_k|\gamma_k\right]$ on both sides of \eqref{eq:lemma-2-1} yields:
		\begin{align}
			&\mathbb{E}\left[Q_k - \gamma_k L_k + \frac{1}{2}m_k -\mu_k^2|\gamma_k\right] \nonumber\\
			\leq &\gamma^{\star} \overline{L}^{\star} -\gamma_k \overline{L}^{\star} + \left(\gamma_k - \gamma^{\star}\right)\mathbb{E}\left[L_k|\gamma_k\right]\nonumber\\
			=&\left(\gamma_k - \gamma^{\star}\right)\mathbb{E}\left[L_k - \overline{L}^{\star}|\gamma_k\right].
		\end{align}
		This finished the proof of the second inequality.
	\end{IEEEproof}
	
	\section{Proof for Lemma \ref{lemma:bound-for-gamma}\label{sec:bound-for-gamma}}
	\begin{IEEEproof}
		To find the upper bound of $\mathbb{E}\left[\sum_{k=1}^{K}\left(F_k - \left(\gamma^{\star} +\overline{D^{\rm v}} + \overline{D^F}\right)L_k\right)\right],$ we first add $\mathbb{E}\left[L_{k-1}D_{k,1}^F+\frac{1}{2}(D^{\rm v}_k)^2+D^{\rm v}_k\max\{D^{\rm a}, \gamma_k\}|\gamma_k\right]$ on both sides of the second inequality of Lemma \ref{lemma:bound-for-Q-L}. Rearranging the terms, we have
		\begin{align}
			&\mathbb{E}\left[Q_k-\gamma^{\star}L_k + \frac{1}{2}m_k - \mu_k^2|\gamma_k\right] \nonumber\\
			&+ \mathbb{E}\left[L_{k-1}D^F_{k,1}+\frac{1}{2}(D^{\rm v}_k)^2+D^{\rm v}_k\max\{D^{\rm a}, \gamma_k\}|\gamma_k\right] \nonumber\\
			=&\mathbb{E}\left[Q_k+L_{k-1}D^F_{k,1}+\frac{1}{2}(D^{\rm v}_k)^2+D^{\rm v}_k\max\{D^{\rm a}, \gamma_k\}|\gamma_k\right]\nonumber\\
			&-\mathbb{E}\left[\gamma^{\star}L_k - \frac{1}{2}m_k + \mu_k^2|\gamma_k\right]
			\nonumber\\
			\leq & -\left(\gamma^{\star} - \gamma_k\right)\left(\mathbb{E}\left[L_k|\gamma_k\right]-\overline{L}^{\star}\right)\nonumber\\
			&+\mathbb{E}\left[L_{k-1}D^F_{k,1}|\gamma_k\right] +\frac{1}{2}\mathbb{E}[m_k]+\overline{D^{\rm v}}\mathbb{E}\left[\max\{D^{\rm a}_k, \gamma_k\}| \gamma_k\right]\nonumber\\
			\overset{(a)}{\leq} &\left(\gamma_k - \gamma^{\star}\right)^2 + L_{k-1}\overline{D^F}+\frac{1}{2}\mathbb{E}[m_k]+\overline{D^{\rm v}}\mathbb{E}\left[\max\{D^{\rm a}_k, \gamma_k| \gamma_k\}\right],
			\label{eq:lemma-3-inequality}
		\end{align}
		where inequality (a) is because $L_{k-1}$ is independent of $D^F_{k,1}$ and $\mathbb{E}\left[L_k|\gamma_k\right]-\overline{L}^{\star}\leq \vert \gamma_k-\gamma^{\star} \vert$.
		
		Deducting $\mathbb{E}\left[\frac{1}{2}m_k-\mu_k^2\right]+\overline{D^F}L_k+\overline{D^{\rm v}_{k-1}}L_k$ from both sides of \eqref{eq:lemma-3-inequality}, we have
		\begin{align}
			=&\mathbb{E}\left[Q_k+L_{k-1}D^F_{k,1}+\frac{1}{2}(D^{\rm v}_k)^2+D^{\rm v}_k\max\{D^{\rm a}, \gamma_k\}|\gamma_k\right]\nonumber\\
			&-\mathbb{E}\left[\gamma^{\star}L_k - \frac{1}{2}m_k + \mu_k^2|\gamma_k\right]
			\nonumber\\
			& - \left(\mathbb{E}\left[\frac{1}{2}m_k-\mu_k^2\right]+\overline{D^F}L_k+\overline{D^{\rm v}_{k-1}}L_k\right)\nonumber\\
			\overset{(a)}{\leq} &\left(\gamma_k - \gamma^{\star}\right)^2 + \left(L_{k-1}-L_k\right)\overline{D^F}\nonumber\\
			&+\overline{D^{\rm v}}\mathbb{E}\left[\max\{D^{\rm a}_k, \gamma_k\}| \gamma_k\right]- \overline{D^{\rm v}_{k-1}}L_k\nonumber\\
			\overset{(b)}{\leq} &\left(\gamma_k - \gamma^{\star}\right)^2 + \left(L_{k-1}-L_k\right)\overline{D^F},
			\label{eq:lemma-3-inequality-1}
		\end{align}
		where inequality (a) is because $\mathbb{E}\left[\gamma^{\star}L_k - \frac{1}{2}m_k + \mu_k^2|\gamma_k\right] = \mathbb{E}\left[\gamma^{\star}L_k - \frac{1}{2}m_k + \mu_k^2\right]$ and inequality (b) is because $\mathbb{E}\left[\max\{D^{\rm a}_k, \gamma_k\}| \gamma_k\right] \leq L_k$.
		
		Therefore, we obtain:
		\begin{align}
			=&\mathbb{E}\left[Q_k+L_{k-1}D^F_{k,1}+\frac{1}{2}(D^{\rm v}_k)^2+D^{\rm v}_k\max\{D^{\rm a}, \gamma_k\}|\gamma_k\right]\nonumber\\
			&-\mathbb{E}\left[\left(\gamma^{\star}+D^F+\overline{D^{\rm v}_{k-1}}\right)L_k\right]\nonumber\\
			\leq &\left(\gamma_k - \gamma^{\star}\right)^2 + \left(L_{k-1}-L_k\right)\overline{D^F}.
			\label{eq:lemma-3-inequality-2}
		\end{align}
		
		Summing \eqref{eq:lemma-3-inequality-2} over epoch $k = 1,2, \cdots, K$, and take the expectation with respect to $\gamma_k$, we complete the proof of Lemma \ref{lemma:bound-for-gamma}:
		\begin{align}
			&\mathbb{E}\left[\sum_{k=1}^{K}\left(Q_k+L_{k-1}D^F_{k,1}+\frac{1}{2}(D^{\rm v}_k)^2+D^{\rm v}_k\max\{D^{\rm a}, \gamma_k\}\right)\right]\nonumber\\
			&-\mathbb{E}\left[\sum_{k=1}^{K}\left(\gamma^{\star}+D^F+\overline{D^{\rm v}_{k-1}}\right)\right] \nonumber\\
			\leq &\mathbb{E}\left[\sum_{k=1}^{K}(\gamma_k \!-\! \gamma^{\star})^2\right] \!-\!\mathbb{E}[L_K]\overline{D^F}
			\!\leq\! \mathbb{E}\left[\sum_{k=1}^{K}\left(\gamma_k \!-\! \gamma^{\star}\right)^2\right]\!.
			\label{eq:lemma-3-1}
		\end{align}
	\end{IEEEproof}
	
	\section{Proof for Lemma \ref{lemma:lower-bound-gamma}\label{appd:proof-lower-bound-gamma}}
	\begin{IEEEproof}
		The proof is divided into two steps. First, we will show that $\gamma_2$ is greater than $\gamma_1$. Then, we will utilize the Taylor expansion to derive the lower bound for $\gamma_2$.

		\textbf{Step 1 ($\gamma_2>\gamma_1$)}:  Define functions 
		$$h_1(\gamma) = \frac{1}{2}\mathbb{E}_{\mathbb{P}_1}\left[\max\{D^{\rm a}, \gamma\}^2\right]
		\!-\!\gamma\mathbb{E}_{\mathbb{P}_1}\left[\max\{D^{\rm a}, \gamma\}\!+\!D^{\rm v}\right] \!+\! N_1,$$
		$$h_2(\gamma) = \frac{1}{2}\mathbb{E}_{\mathbb{P}_2}\left[\max\{D^{\rm a}, \gamma\}^2\right]
		\!-\!\gamma\mathbb{E}_{\mathbb{P}_2}\left[\max\{D^{\rm a}, \gamma\}\!+\!D^{\rm v}\right] \!+\! N_2,$$
		where $N_1=\frac{1}{2}\mathbb{E}_{\mathbb{P}_1}[{D^{\rm v}}^2]-\mathbb{E}_{\mathbb{P}_1}[D^{\rm v}]^2$ and $N_2=\frac{1}{2}\mathbb{E}_{\mathbb{P}_2}[{D^{\rm v}}^2]-\mathbb{E}_{\mathbb{P}_2}[D^{\rm v}]^2$
		By the definition of $\gamma_1$ and $\gamma_2$, we have $h_1(\gamma_1)=0$ and $h_2(\gamma_2)=0$. Furthermore, the function $h_2(\gamma)$ is monotonically decreasing, as validated through the derivative of $h_2(\gamma)$ \cite{pan2023ageoptimal}:
		\begin{equation}
			h_2(\gamma)^{\prime} = -\mathbb{E}_{\mathcal{P}_2}\!\left[\max\{D^{\rm a}, \gamma\}+D^{\rm v}\right] < 0.
		\end{equation}
		
		Since $h_2(\gamma_2) = 0$ and that $h_2(\gamma)$ is monotonically decreasing, we will prove $\gamma_2>\gamma_1$ by showing $h_2(\gamma_1)>0$. In addition, because $h_1(\gamma_1) =0$, it's sufficient to show that $h_2(\gamma_1)>h_1(\gamma_1)$. Let $\mathbb{P}_{1, D^{\rm v}}, \mathbb{P}_{2, D^{\rm v}}$ be the distributions of $D^{\rm v}$ when the packet transmission failure probablility is $\alpha_1, \alpha_2$, respectively. Then we can compute $h_2(\gamma)-h_1(\gamma)$ as follows:
		\begin{align}
			&h_2(\gamma)-h_1(\gamma)\nonumber\\
			=
			&\gamma\left(\mathbb{E}_{\mathbb{P}_{1, D^{\rm v}}}[D^{\rm v}]-\mathbb{E}_{\mathbb{P}_{2, D^{\rm v}}}\left[D^{\rm v}\right]\right)+N_1-N_2\nonumber\\
			=&\frac{1}{2}\left(\mathbb{E}_{\mathbb{P}_{2, D^{\rm v}}}\left[\left(D^{\rm v}\right)^2\right]-\mathbb{E}_{\mathbb{P}_{1, D^{\rm v}}}\left[\left(D^{\rm v}\right)^2\right]\right)\nonumber\\
			&+\left(\gamma+\mathbb{E}_{\mathbb{P}_{1, D^{\rm v}}}\left[D^{\rm v}\right]+\mathbb{E}_{\mathbb{P}_{2, D^{\rm v}}}\left[D^{\rm v}\right]\right)\nonumber\\
			&\cdot\left(\mathbb{E}_{\mathbb{P}_{1, D^{\rm v}}}\left[D^{\rm v}\right]-\mathbb{E}_{\mathbb{P}_{2, D^{\rm v}}}\left[D^{\rm v}\right]\right).
			\label{eq:h1h2-1}
		\end{align}
		
		Since the channel reliability follows a Bernoulli distribution with parameter $\alpha$, the number of transmission attempts $M_k$ in each epoch follows a geometric distribution. Consequently, the expected values of $D^{\rm v}$ with $\alpha_i$ under distribution $\mathbb{P}_{i, D^{\rm v}}$ can be expressed as $\mathbb{E}_{\mathbb{P}_{i, D^{\rm v}}}[D^{\rm v}]=(\frac{1}{1-\alpha_i}-1)\left(\mathbb{E}[D^F]+\mathbb{E}[D^B]\right), \mathbb{E}_{\mathbb{P}_{i, D^{\rm v}}}[(D^{\rm v})^2]=\frac{\alpha_i(\alpha_i+1)}{(1-\alpha_i)^2}\mathbb{E}\left[\left(D^{\rm a}\right)^2\right]$. Then we can further simplify \eqref{eq:h1h2-1} as follows:
		\begin{align}
			&h_2(\gamma)-h_1(\gamma)\nonumber\\
			=&\frac{1}{2}\left(\frac{\alpha_2(\alpha_2+1)}{(1-\alpha_2)^2}-\frac{\alpha_1(\alpha_1+1)}{(1-\alpha_1)^2}\right)\mathbb{E}\left[\left(D^{\rm a}\right)^2\right]\nonumber\\
			&+\left(\gamma+\left(\frac{\alpha_1}{1-\alpha_1}+\frac{\alpha_2}{1-\alpha_2}\right)\mathbb{E}\left[D^{\rm a}\right]\right)\nonumber\\
			&\cdot\left(\frac{1}{1-\alpha_1}-\frac{1}{1-\alpha_2}\right)\mathbb{E}\left[D^{\rm a}\right]\nonumber\\
			=&\frac{1}{2}\frac{(\alpha_2-\alpha_1)(\alpha_2+\alpha_1+1-3\alpha_1\alpha_2)}{(1-\alpha_2)^2(1-\alpha_1)^2}\mathbb{E}\left[\left(D^{\rm a}\right)^2\right]\nonumber\\
			&+\left(\gamma+\frac{\alpha_1+\alpha_2-2\alpha_1\alpha_2}{(1-\alpha_1)(1-\alpha_2)}\mathbb{E}\left[D^{\rm a}\right]\right)\nonumber\\
			&\cdot\frac{\alpha_1-\alpha_2}{(1-\alpha_1)(1-\alpha_2)}\mathbb{E}\left[D^{\rm a}\right]\nonumber\\
			=&\frac{\alpha_1-\alpha_2}{(1-\alpha_1)^2(1-\alpha_2)^2}\Big((\alpha_1+\alpha_2-2\alpha_1\alpha_2)\mathbb{E}\left[D^{\rm a}\right]^2\nonumber\\
			&+\gamma(1-\alpha_1)(1-\alpha_2)\mathbb{E}\left[D^{\rm a}\right]\nonumber\\
			&-\frac{1}{2}(\alpha_1+\alpha_2+1-3\alpha_1\alpha_2)\mathbb{E}\left[\left(D^{\rm a}\right)^2\right]\Big).
			\label{eq:h2h1-diff}
		\end{align}
		
		Recall that the delay distributions follow uniform distribution, i.e., $\mathbb{P}_{\text{FD}}=\mathbb{P}_{\text{BD}}=\text{Uni}[0,1]$, we have $\mathbb{E}\left[\left(D^{\rm a}\right)^2\right]=\frac{7}{6}, \mathbb{E}\left[D^{\rm a}\right]^2=1$. With $\alpha_1=0.5$, we can obtain the optimal threshold for $\mathbb{P}_1$: $\gamma_1 = 1.6759$ by numerical calculation. Then, for $\alpha_2 <\alpha_1 = 0.5$, we have $h_2(\gamma_1)-h_1(\gamma_1)>0$ and therefore $h_2(\gamma_1) > h_1(\gamma_1) = 0$. Since $h_2(\gamma_2)=0$ and $h_2(\gamma)$ is monotonically decreasing, we can conclude that $\gamma_2 > \gamma_1$.
		
		\textbf{Step 2 (Taylor expansion):} We will continue to give the lower bound of $\gamma_2-\gamma_1$ through Taylor expansion of $h_2(\cdot)$. By Taylor expansion, we have
		\begin{equation}
			\gamma_2 - \gamma_1 = \frac{h_2(\gamma_2)-h_2(\gamma_1)}{h_2^{\prime}(\gamma)}=\frac{h_2(\gamma_1)-h_2(\gamma_2)}{-h_2^{\prime}(\gamma)},
			\label{eq:taylor-exp}
		\end{equation}
		where $\gamma\in [\gamma_1,\gamma_2]$. We proceed by giving the lower bound of $h_2(\gamma_1)-h_2(\gamma_2)$ and the upper bound of $h_2^{\prime}(\gamma)$. For $h_2^{\prime}(\gamma)$, we will first bound $\gamma_2$ and then give the upper bound. According to \eqref{eq:gamma-upp}, as $\alpha_2<0.5$, we can upper bound $\gamma_2$ by
		\begin{align}
			\gamma_2\leq& \frac{\frac{1}{2}\overline{H}}{\overline{D}}-\overline{D^{\rm v}}\nonumber\\
			\leq&\frac{\frac{1}{2}\cdot \left(\frac{7}{6}+\frac{2\times0.5}{1-0.5}+\frac{7}{6}\times\frac{0.5\times(1+0.5)}{(1-0.5)^2}\right)}{2}
			<  2.
		\end{align}
		Therefore, the derivative of $h_2(\gamma)$ can be upper bounded by:
		\begin{align}
			|h_2^{\prime}(\gamma)|=&\mathbb{E}_{\mathbb{P}_2}\left[\left(\gamma-D^{\rm a}\right)^{+}+D^{\rm a}+D^{\rm v}\right]\nonumber\\
			\leq&\gamma_2+\mathbb{E}_{\mathbb{P}_2}\left[D^{\rm a}+D^{\rm v}\right]
			\leq2+5
			=7.
			\label{eq:bound-h-deriva}
		\end{align}
		
		For the lower bound of $h_2(\gamma_1)-h_2(\gamma_2)$, notice that $h_2(\gamma_2)=0$ and $h_1(\gamma_1)=0$, lower bounding $h_2(\gamma_1)-h_2(\gamma_2)$ is equivalent to lower bounding $h_2(\gamma_1)-h_1(\gamma_1)$. Use the result from Equation \eqref{eq:h2h1-diff} in \textbf{Step 1}, and recall that $\alpha_1-\alpha_2 = \frac{1}{4\sqrt{k}}$, we have:
		\begin{align}
			&h_2(\gamma_1)-h_1(\gamma_1)\nonumber\\
			=&\frac{\alpha_2-\alpha_1}{(1-\alpha_1)^2(1-\alpha_2)^2}\Big(\frac{1}{2}(\alpha_1+\alpha_2+1-3\alpha_1\alpha_2)\nonumber\\
			&\cdot\mathbb{E}\left[\left(D^{\rm a}\right)^2\right]-\gamma(1-\alpha_1)(1-\alpha_2)\mathbb{E}\left[D^{\rm a}\right]^2\nonumber\\
			&-(\alpha_1+\alpha_2-2\alpha_1\alpha_2)\mathbb{E}\left[D^{\rm a}\right]^2\Big)
			=\frac{1}{\sqrt{k}}N_1.
			\label{eq:bound-h-function}
		\end{align}
		
		Plugging the upper bound of $|h_2^{\prime}(\gamma)|$, i.e., inequality \eqref{eq:bound-h-deriva} and the lower bound of $h_2(\gamma_1)-h_1(\gamma_1)$, i.e., equality \eqref{eq:bound-h-function} into the Taylor expansion expression \eqref{eq:taylor-exp}, we can lower bound $\gamma_2-\gamma_1$ by
		\begin{equation}
			\gamma_2-\gamma_1\geq \frac{N_1}{7}\frac{1}{\sqrt{k}}.
			\label{eq:diff-gamma}
		\end{equation}
	\end{IEEEproof}
	
	\section{Proof for Lemma \ref{lemm:bound-for-reward}\label{appd:relation-reward}}
	\begin{IEEEproof}	
		Denote  $\Pi_l \triangleq \{w\vert \mathbb{E}[D^{\rm a}+w+D^{\rm v}]=l, \forall \text{ stationary policy } w\}$ to be the set of stationary policies whose expected cycle length is $l$. If $L$ satisfies $\overline{D^{\rm a}}+\overline{D^{\rm v}} \leq l \leq \overline{D^{\rm a}}+\overline{D^{\rm v}}+W_{\text{ub}}$, then the set will not be empty. Next, we will establish a lower bound for the expected reward $q$, which can be formulated into an optimization problem:
		\begin{pb}
			\begin{align}
				q_{\text{l, opt}} \triangleq &\inf_{w} \mathbb{E}\left[\frac{1}{2}\left(D^{\rm a}+w+D^{\rm v}\right)^2\right], \nonumber\\ 
				& \text{s.t. } \mathbb{E}\left[D^{\rm a}+w+D^{\rm v}\right]=l.
			\end{align}
			\label{pb:l-length-reward}
		\end{pb}
		
		Problem \ref{pb:l-length-reward} can be solved through Lagrange multiplier approach. The function is as follows:
		\begin{align}
			\mathcal{L}_1(w, &\lambda, \mu)\triangleq\frac{1}{2}\mathbb{E}\left[\left(D^{\rm a}+w+D^{\rm v}\right)^2\right]\nonumber\\
			&+\lambda\left(\mathbb{E}\left[D^{\rm a}+w+D^{\rm v}\right]-l\right)+\mathbb{E}[w \mu],
		\end{align}
		where $\lambda$ and $\mu=\mu(D^{\rm a})\geq 0$ are dual variables. For function $\theta(\cdot) \in L_3$, the Gateaux derivative of the Lagrange function is denoted by $\delta\mathcal{L}_1(w;\lambda, \mu, \theta)$:
		\begin{align}
			\delta\mathcal{L}_1(w;\lambda, \mu, \theta)&\!=\!\lim_{\epsilon \rightarrow 0}\frac{\delta\mathcal{L}_1(w+\epsilon\theta,\lambda, \mu) - \mathcal{L}(w, \lambda, \mu)}{\epsilon}\nonumber\\
			&\!=\!\mathbb{E}\left[\left(D^{\rm a}+w+\lambda+\mu+D^{\rm v}\right)\theta\right].
			\label{eq:gate-deriva}
		\end{align}
		
		\begin{subequations}
			The primal feasibility of the Karush-Kuhn-Tucker (KKT) condition requires:
			\begin{equation}
				\delta\mathcal{L}_1(w;\lambda, \mu, \theta) = 0, \forall \theta \in L_3,
				\label{eq:KKT-condition}
			\end{equation}
			and the Complete Slackness (CS) conditions require the Lagrange multipliers corresponding to the equality constraints are zero, i.e.,
			\begin{align}
				&\lambda\left(\mathbb{E}\left[D^{\rm a}+w+D^{\rm v}\right]-L\right) = 0,\label{eq:CS-1}\\
				&w \mu  = 0, \hspace{0.5cm}\forall D^{\rm a}.\label{eq:CS-2}
			\end{align}
		\end{subequations}
		
		Plugging Gateaux derivative \eqref{eq:gate-deriva} into the KKT conditions \eqref{eq:KKT-condition} and considering the CS conditions \eqref{eq:CS-1} and \eqref{eq:CS-2}, the optimal policy $w^{\star}_L$ to Problem \ref{pb:l-length-reward} can be obtained as follows:
		\begin{equation}
			w^{\star}_l(D^{\rm a}) = (\gamma_l - D^{\rm a})^{+},
		\end{equation}
		where the threshold $\gamma_l$ satisfies:
		\begin{equation}
			\mathbb{E}\left[\left(\gamma_l-D^{\rm a}\right)^{+}\right] = l-\overline{D}.
		\end{equation}
		
		Before lower bounding the reward $q$, i.e., $\mathbb{E}\left[\frac{1}{2}\left((\gamma_l-D^{\rm a})^{+}+D^{\rm a}+D^{\rm v}\right)^2\right]$, we provide the connection between the difference of thresholds with the epoch lengths. Recall that $\gamma^{\star}$ is the optimal updating threshold and leads to an average epoch length of
		\begin{equation}
			\overline{L}^{\star} = \mathbb{E}\left[D^{\rm a}+(\gamma^{\star}-D^{\rm a})^{+}+D^{\rm v}\right].
		\end{equation}
		Under the same distribution set $\mathbb{P}$, for any threshold $\gamma_1\geq \gamma_2$, the waiting time under $\gamma_1$, i.e., $(\gamma_1-D^{\rm a})^{+}$ will always be greater than the waiting time under $\gamma_2$, i.e., $(\gamma_2-D^{\rm a})^{+}$. Therefore, the epoch length under $\gamma_1$ and $\gamma_2$ satisfy:
		\begin{align}
			0 \leq & \mathbb{E}\left[\left(\gamma_1-D^{\rm a}\right)^{+}+D^{\rm a}+D^{\rm v}\right]\nonumber\\
			&-\mathbb{E}\left[\left(\gamma_2-D^{\rm a}\right)^{+}+D^{\rm a}+D^{\rm v}\right]\nonumber\\
			=&\mathbb{E}\left[\left(\gamma_1-\gamma_2\right)\mathbb{I}\left(D\leq \gamma_1\right)\right]\nonumber\\
			&+\mathbb{E}\left[\left(\gamma_1-D^{\rm a}\right)\mathbb{I}\left(\gamma_2\leq D^{\rm a}\leq \gamma_1\right)\right]\nonumber\\
			\leq& \gamma_1-\gamma_2.
			\label{eq:gamma-difference-inequality}
		\end{align}
		
		Using inequality \eqref{eq:gamma-difference-inequality}, the difference between $\gamma_l$ and $\gamma^{\star}$ can be lower bounded by the difference of the epoch length:
		\begin{equation}
			\vert\gamma_l-\gamma^{\star}\vert \geq \vert l-\overline{L}^{\star}\vert.
			\label{eq:gamma-epoch-rela}
		\end{equation}
		
		Finally, we proceed to establish the lower bound for $\mathbb{E}\left[\frac{1}{2}\left(\left(\gamma_l-D^{\rm a}\right)^{+}+D^{\rm a}+D^{\rm v}\right)^2\right]$ by considering the following two cases.
		\begin{enumerate}
			\item \textbf{Case 1}: $l\geq \overline{L}^{\star}$, it can be easily verify that $\gamma_l \geq \gamma^{\star}$. Therefore, we have
			\begin{align}
				&\frac{1}{2}\mathbb{E}\left[\left((\gamma_l-D^{\rm a})^{+}+D^{\rm a}+D^{\rm v}\right)^2\right]\nonumber\\
				=&\frac{1}{2}\mathbb{E}\left[(\gamma_l+D^{\rm v})^2\mathbb{I}(D^{\rm a}\leq \gamma_l)\right]\nonumber\\
				&+\frac{1}{2}\mathbb{E}\left[\left(D^{\rm a}\right)^2\mathbb{I}(D^{\rm a}\geq \gamma_l)\right]\nonumber\\
				=&\frac{1}{2}\mathbb{E}\left[(\gamma^{\star}+D^{\rm v})^2\mathbb{I}(D^{\rm a}\leq \gamma^{\star})\right]\nonumber\\
				&+\frac{1}{2}\mathbb{E}\left[\left(D^{\rm a}+D^{\rm v}\right)^2\mathbb{I}(D^{\rm a}\geq \gamma^{\star})\right]\nonumber\\
				&+\frac{1}{2}\mathbb{E}\left[\left(\left(\gamma_l+D^{\rm v}\right)^2\!-\!\left(\gamma^{\star}+D^{\rm v}\right)^2\right)\mathbb{I}(D^{\rm a}\leq \gamma^{\star})\right]\nonumber\\
				&+\frac{1}{2}\mathbb{E}\Big[\left(\left(\gamma_l+D^{\rm v}\right)^2-\left(D^{\rm a}+D^{\rm v}\right)^2\right)\cdot \nonumber\\
				&\quad \mathbb{I}(\gamma^{\star}\leq D^{\rm a}\leq \gamma_l)\Big]\nonumber\\
				\overset{(a)}{\geq} &\overline{Q}^{\star}+\frac{1}{2}\mathbb{E}\left[\left(\gamma_l-\gamma^{\star}\right)^2\mathbb{I}(D^{\rm a} \leq \gamma^{\star})\right]\nonumber\\
				&+\mathbb{E}\left[\gamma^{\star}(\gamma_l-\gamma^{\star})\mathbb{I}(D^{\rm a}\leq \gamma^{\star})\right]\nonumber\\
				&+\mathbb{E}\left[D^{\rm v}(\gamma_l-\gamma^{\star})\mathbb{I}(D^{\rm a}\leq \gamma^{\star})\right]\nonumber\\
				&+\mathbb{E}\left[\gamma^{\star}\left(\gamma_l-D^{\rm a}\right)\mathbb{I}(\gamma^{\star}\leq D^{\rm a}\leq \gamma_l)\right]\nonumber\\
				&+\mathbb{E}\left[D^{\rm v}(\gamma_l-D^{\rm a})\mathbb{I}(\gamma^{\star}\leq D^{\rm a}\leq \gamma_l)\right]\nonumber\\
				\overset{(b)}{\geq} &\gamma^{\star}\overline{L}^{\star}+\mathbb{E}\le[D^{\rm v}]\overline{L}^{\star} + \frac{1}{2}p_w(\gamma_l-\gamma^{\star})^2\nonumber\\
				&+\left(\gamma^{\star}+D^{\rm v}\right)(L-\overline{L}^{\star})\nonumber\\
				\overset{(c)}{\geq} & \left(\gamma^{\star}+D^{\rm v}\right)l+ \frac{1}{2}p_w(L-\overline{L}^{\star})^2.
				\label{eq:lemma-4-1}
			\end{align}
			Inequality (a) is from inequality $\gamma_l^2-(\gamma^{\star})^2 \geq (\gamma_l-\gamma^{\star})^2+2\gamma^{\star}(\gamma_l-\gamma^{\star})$ and for delays satisfy $\gamma^{\star} \leq D^{\rm a} \leq \gamma_l^{\star}$, we have $(\gamma_l^{\star})^2-(D^{\rm a})^2 = (D^{\rm a})(\gamma_l^{\star}-D^F-D^B) \geq \gamma^{\star}(\gamma^{\star}_L-D^{\rm a})$. Inequality (b) is true by considering the difference of epoch length $L-\overline{L}^{\star}$ as the sum of two expectations: $=\mathbb{E}[(\gamma_l-\gamma^{\star})\mathbb{I}(D^{\rm a}\leq \gamma^{\star})]+\mathbb{E}[(\gamma_l-D^{\rm a})\mathbb{I}(\gamma^{\star} \leq D^{\rm a}\leq \gamma_l)]$ Since the delays $D^F, D^B, D^{\rm v}$ are independent, the expectation can be simplified and we obtain the inequality. Inequality (c) is because the upper bound of $\gamma_l-\gamma^{\star}$ previously stated in \eqref{eq:gamma-epoch-rela}.
			\item \textbf{Case 2:} $l\leq \overline{L}
			^{\star}$, similarly, it can be verified that $\gamma_l \leq \gamma^{\star}$. As a result, we have
			\begin{align}
				&\frac{1}{2}\mathbb{E}\left[\left((\gamma_l-D^{\rm a})^{+}+D^{\rm a}+D^{\rm v}\right)^2\right]\nonumber\\
				=&\frac{1}{2}\mathbb{E}\left[\left(\gamma_l+D^{\rm v}\right)^2\mathbb{I}(D^{\rm a}\leq \gamma_l)\right]\nonumber\\
				&+\frac{1}{2}\mathbb{E}\left[\left(D^{\rm a}+D^{\rm v}\right)^2\mathbb{I}(D^{\rm a}\geq \gamma_l)\right]\nonumber\\
				=&\frac{1}{2}\mathbb{E}\left[(\gamma^{\star}+D^{\rm v})^2\mathbb{I}(D^{\rm a}\leq \gamma^{\star})\right]\nonumber\\
				&+\frac{1}{2}\mathbb{E}\left[\left(D^{\rm a}+D^{\rm v}\right)^2\mathbb{I}(D^{\rm a}\geq \gamma^{\star})\right]\nonumber\\
				&+\frac{1}{2}\mathbb{E}\left[\left(\left(\gamma_l+D^{\rm v}\right)^2-\left(\gamma^{\star}+D^{\rm v}\right)^2\right)\mathbb{I}(D^{\rm a}\leq \gamma^{\star})\right]\nonumber\\
				&-\frac{1}{2}\mathbb{E}\Big[\left(\left(\gamma_l+D^{\rm v}\right)^2-\left(D^{\rm a}+D^{\rm v}\right)^2\right)\cdot \nonumber\\
				&\quad \mathbb{I}(\gamma_l\leq D^{\rm a}\leq \gamma^{\star})\Big]\nonumber\\
				\overset{(d)}{\geq} &\overline{Q}^{\star}+\frac{1}{2}\mathbb{E}\left[\left(\gamma_l-\gamma^{\star}\right)^2\mathbb{I}(D^{\rm a} \leq \gamma^{\star})\right]\nonumber\\
				&+\mathbb{E}\left[\gamma^{\star}(\gamma_l-\gamma^{\star})\mathbb{I}(D^{\rm a}\leq \gamma^{\star})\right]\nonumber\\
				&+\mathbb{E}\left[D^{\rm v}(\gamma_l-\gamma^{\star})\mathbb{I}(D^{\rm a}\leq \gamma^{\star})\right]\nonumber\\
				&-\mathbb{E}\left[\gamma^{\star}\left(\gamma_l-D^{\rm a}\right)\mathbb{I}(\gamma_l\leq D^{\rm a}\leq \gamma^{\star})\right]\nonumber\\
				&-\mathbb{E}\left[D^{\rm v}(\gamma_l-D^{\rm a})\mathbb{I}(\gamma_l\leq D^{\rm a}\leq \gamma^{\star})\right]\nonumber\\
				\overset{(e)}{\geq} &\gamma^{\star}\overline{L}^{\star}\!+\!\mathbb{E}[D^{\rm v}]\overline{L}^{\star}\! +\! \frac{1}{2}p_w(\gamma_l-\gamma^{\star})^2\!+\!\left(\gamma^{\star}+D^{\rm v}\right)(l-\overline{L}^{\star})\nonumber\\
				\overset{(f)}{\geq} & \left(\gamma^{\star}+D^{\rm v}\right)l + \frac{1}{2}p_w(l-\overline{L}^{\star})^2.
				\label{eq:lemma-4-2}
			\end{align}
		\end{enumerate}
		Inequality (d), (e), and (f) are similar to inequality (a), (b), and (c).
		
		Combining the result in \eqref{eq:lemma-4-1} and \eqref{eq:lemma-4-2}, we can obtain the statement in Lemma 4:
		\begin{equation}
			q_l \geq \left(\gamma^{\star}+D^{\rm v}\right)l + \frac{1}{2}p_w(l-\overline{L}^{\star})^2.
		\end{equation}
		
	\end{IEEEproof}

	\section{Proof for Lemma \ref{lemm:bound-for-z}\label{appd:proof-bound-for-z}}
	\begin{IEEEproof}
		The minimax risk bound on $\hat{l}-\overline{L}^{\star}$ is established similarly using the Le Cam's two-point method. Let $\mathbb{P}_1$ and $\mathbb{P}_2$ be two distribution sets defined in Appendix \ref{appd:converse-bound-1}. Denote $L_1=\mathbb{E}_{\mathbb{P}_1}[(\gamma_1-D^{\rm a})^{+}+D^{\rm a}+D^{\rm v}]$ and $L_2 = \mathbb{E}_{\mathbb{P}_2}[(\gamma_2-D^{\rm a})^{+}+D^{\rm a}+D^{\rm v}]$ be the optimal epoch length by using AoI minimum policies $w^{\star}_{\mathbb{P}_1}$ and $w^{\star}_{\mathbb{P}_2}$. By Le Cam's inequality, we have:
		\begin{align}
			\inf_{\hat{l}}\sup_{\mathbb{P}}\mathbb{E}\left[\left(\hat{l}(\mathcal{H}_k)\!-\!\overline{L}^\star(\mathbb{P})\right)^2\right]\geq(L_1\!-\!L_2)^2\cdot\mathbb{P}_1^{\otimes k}\!\wedge\!\mathbb{P}_2^{\otimes k}.
			\label{eq:le-cam-ineq}
		\end{align}
		
		Similar to the proof in Appendix \ref{appd:converse-bound-1}, to use Le Cam's two-point method, we need to lower bound $L_2-L_1$ and $\mathbb{P}_1^{\otimes k}\wedge\mathbb{P}_2^{\otimes k}$ respectively. The lower bound on $\mathbb{P}_1^{\otimes k}\wedge\mathbb{P}_2^{\otimes k}$ can be obtained in (55) and the lower bound on $L_2-L_1$ can be obtained as follows:
		\begin{align}
			L_2-L_1
			=&\mathbb{E}_{\mathbb{P}_{\text{FD}}, \mathbb{P}_{\text{BD}}, \mathbb{P}_{2, D^{\rm v}}}[(\gamma_2-D^{\rm a})^{+}+D^{\rm a}+D^{\rm v}]\nonumber\\
			&-\mathbb{E}_{\mathbb{P}_{\text{FD}}, \mathbb{P}_{\text{BD}}, \mathbb{P}_{1, D^{\rm v}}}[(\gamma_1-D^{\rm a})^{+}+D^{\rm a}+D^{\rm v}]\nonumber\\
			\overset{(a)}{\geq}&\mathbb{E}_{\mathbb{P}_{2, D^{\rm v}}}[D^{\rm v}]-\mathbb{E}_{\mathbb{P}_{1, D^{\rm v}}}[D^{\rm v}]\nonumber\\
			=&\frac{\alpha_1-\alpha_2}{(1-\alpha_1)(1-\alpha_2)}\mathbb{E}[D^{\rm a}]
			\overset{(b)}{\geq}  \frac{1}{\sqrt{k}} N_3,
			\label{eq:diff-epoch-length}
		\end{align}
		where inequality (a) is because for $\gamma_2>\gamma_1$ and inequality (b) is obtained from \eqref{eq:diff-gamma} and $N_3=\frac{1}{(1-\alpha_1)(1-\alpha_2)}$.
		
		Plugging the lower bound of $l_2-l_1$ \eqref{eq:diff-epoch-length} into the Le Cam's inequality \eqref{eq:le-cam-ineq}, we finish the proof for Lemma 5:
		\begin{equation}
			\inf_{\hat{l}}\sup_{\mathbb{P}}\mathbb{E}[(\hat{l}(\mathcal{H}_k)-\overline{L}^{\star}(\mathbb{P}))^2]\geq\frac{N_3^2}{49}\left(1-\frac{\sqrt{N_2}}{4}\right)^2\cdot\frac1k=\Omega(\frac{1}{k}).
		\end{equation}
	\end{IEEEproof}
 	\end{document}